\documentclass[sigconf]{acmart}

\setcopyright{none}
\renewcommand\footnotetextcopyrightpermission[1]{}

\usepackage{tikz}
\usepackage{amsmath}
\usepackage{paralist}
\usepackage{enumitem}

\usepackage{caption}
\usepackage[section]{placeins}
\usepackage{array}
\usepackage{booktabs}
\usepackage{longtable}
\usepackage{stfloats}
\usepackage{subcaption}
\usepackage{tabularx}
\usepackage{tabulary}
\usepackage{xurl}
\usepackage{hyperref}
\usepackage{float}
\usepackage{xspace}
\usepackage{multirow}
\usepackage{enumitem}
\usepackage{listings}
\usepackage{mdframed}
\usepackage{url}
\usepackage{paralist}

\setlist[itemize]{leftmargin=*}
\setlist[enumerate]{leftmargin=*}

\newcommand{\rv}{reCAPTCHA\xspace}
\newcommand{\rvs}{reCAPTCHAs\xspace}
\newcommand{\rvi}{reCAPTCHA v1\xspace}
\newcommand{\rvis}{reCAPTCHA v1s\xspace}
\newcommand{\rvii}{reCAPTCHAv2\xspace}
\newcommand{\rviis}{reCAPTCHAv2s\xspace}
\newcommand{\rviii}{reCAPTCHAv3\xspace}

\newcommand{\Captchas}{\textsc{Captchas}\xspace}
\newcommand{\captcha}{\textsc{captcha}\xspace}
\newcommand{\captchas}{\textsc{captchas}\xspace}

\newcommand{\taggedpara}[1]{\textbf{#1}\xspace}

\ifdefined\showchanges

\else

\fi

\begin{document}
 \pagestyle{plain}

    \title{Dazed \& Confused: \\
    A Large-Scale Real-World User Study of reCAPTCHAv2}

        \author{Andrew Searles}
        \email{searlesa@uci.edu}
        \affiliation{UC Irvine}
        \author{Renascence Tarafder Prapty}
        \email{rprapty@uci.edu}
        \affiliation{UC Irvine}
        \author{Gene Tsudik}
        \email{gene.tsudik@uci.edu}
        \affiliation{UC Irvine}



		\begin{abstract}
	Since about 2003, \captchas have been widely used as a barrier against 
bots, while simultaneously annoying great multitudes of users worldwide. 
As their use grew, techniques to defeat or bypass \captchas kept improving, while \captchas 
themselves evolved in terms of sophistication and diversity, becoming increasingly difficult to 
solve for both bots and humans. Given this long-standing and still-ongoing arms race, it is 
important to investigate usability, solving performance, and user perceptions of modern 
\captchas. In this work, we do so via a large-scale (over $3,600$ distinct users) 
13-month real-world user study and post-study survey. The study, conducted at a 
large public university, was based on a live account creation and password recovery 
service with currently prevalent \captcha type: \rvii. 

Results show that, with more attempts, users improve in solving checkbox challenges.
For website developers and user study designers, results indicate that the website context 
directly influences (with statistically significant differences) solving time between password 
recovery and account creation. We consider the impact of participants' major and education level, 
showing that certain majors exhibit better performance, while, in general, education level 
has a direct impact on solving time. Unsurprisingly, we discover that participants find 
image challenges to be annoying, while checkbox challenges are perceived as easy. We also 
show that, rated via System Usability Scale (SUS), image tasks are viewed as "OK", while 
checkbox tasks are viewed as "good". 

We explore the cost and security of \rvii and conclude that it has an immense cost and no security. 
Overall, we believe that this study's results prompt a natural conclusion: 
{\em \rvii and similar \rv technology should be deprecated.}

	\end{abstract}
	\maketitle	
	


	\section{Introduction}
\label{sec:intro}
Many types of Internet-based activities and services require verification of human presence, e.g.,
ticket sales, reservations, and account creation. Left unchecked, bots will 
gobble up most resources available through such activities: they are much faster and
way more agile than any human or a group thereof. This problem is not new: the first seminal step to
combat it took place in 2003 when von Ahn et al.~\cite{vonAhn} proposed \captcha as an automated test 
that is supposed to be easy for humans to pass, yet difficult or impossible for computer programs 
(aka bots) at the time. The key conjecture underlying the \captcha concept is that, if a computer 
program successfully solved \captchas, then the same program could be repurposed to solve some 
computationally hard AI problem.

This seemed to be a win-win situation: either \captchas attest to genuine human presence or they 
spur a significant advance in AI technology. Furthermore, \captchas were touted as a tool for the 
common good, since human-based solutions helped with difficult (for computers) and useful tasks, 
such as recognizing blurred text that confounded OCR algorithms, 
or labeling photos with names of objects appearing in them in order to aid image classification.

Another major advance occurred in 2007 when von Ahn et al. introduced \rv~\cite{vonAhn_recaptcha}.
\rv was designed to reuse challenge results as a form of human-based data labeling for advancing machine learning.
Google acquired \rv in late 2009~\cite{google_recaptcha_2009} and, by June 2010, it was reported 
that \rv had over $100$ million distinct daily users~\cite{recap_FAQ_100mil}. Assuming that 
this number stayed constant since 2010 (though it most likely grew significantly), over half a 
trillion \rvs have been solved in the meantime. 
This collectively amounts to an immense human cost.

However, almost from the start, an ``arms race'' began between bot and \captcha developers.
Most early \captcha types were based on recognition of distorted text. Unfortunately, as a consequence 
of rapid advances in machine learning and computer vision, bots evolved to quickly and accurately 
recognize and classify distorted text~\cite{yan2008low,gao2012divide,hernandez2010pitfalls}, 
reaching over 99\% accuracy by 2014~\cite{Goodfellow,Shet}. To this end, in 2012 Google switched from 
distorted text to image classification, using images from the Google Street View project~\cite{perez_2012}.
This transition ended in 2014 with the introduction of \rvii \cite{intro_reCAPTCHAv2}, which uses a 
two-step process: (1) a combination of behavioral analysis and a simple checkbox, and 
(2) image classification tasks as a fallback for users who fail the checkbox challenge~\cite{reCAPTCHAv2}.
By 2016, both (1) and (2) were defeated with a high degree of accuracy by bots~\cite{Sivakorn16}.

Regardless of its diminished efficacy, \rv remains to be the prevalent \captcha type on the 
Internet \cite{cap_usage}, deployed on over $13$ million websites in 2023. It is therefore important 
to periodically evaluate and quantify its impact in terms of usability, solving performance, and 
user perceptions. 

Several prior \captcha user studies explored solving performance, e.g., 
\cite{searlesa_usenix,Bursztein,Bigham,Gao,Ross,Uzun,manarDynamic2014,mohitAutomatic2019,gaoEmerging2019, 
Krol,fidas2011, senCAP, Ho, tanth19}. Also, \cite{manarDynamic2014, gaoEmerging2019, senCAP} 
looked into usability of \captchas via the well-known SUS scale.
\cite{searlesa_usenix,fidas2011, Krol, Bigham, tanth19, Gao, mohitAutomatic2019} studied user preferences 
related to \captcha types. However, only two recent (2019/2023) user studies \cite{tanth19, searlesa_usenix} involved 
\rvii. However, \cite{tanth19} had relatively few participants (40), used unclear methodology, and did 
not consider usability.
\cite{searlesa_usenix} presents interesting comparison points discussed in Section~\ref{sec:related_work}.
Most other user studies \cite{manarDynamic2014,mohitAutomatic2019,gaoEmerging2019, 
fidas2011, Gao, senCAP} were conducted on newly proposed (and therefore, mocked-up) \captcha types.

Furthermore, many previous \captcha studies 
\cite{searlesa_usenix,Bursztein,manarDynamic2014,mohitAutomatic2019,gaoEmerging2019,Ho,senCAP} were conducted on Amazon 
Mechanical Turk (MTurk) \cite{mturk}, which exhibits data quality issues \cite{webb2022too}. 
Also, all these studies involved some bias, since participants were informed about 
study goals, i.e., they were selected based on their willingness to solve \captchas, 
for a certain monetary reward.

The above discussion motivates the work presented in this paper, the centerpiece of which
is a large-scale ($>3,600$ participants) 13-month IRB-appproved user study of 
\rvii. The study was conducted using a live account creation (and password recovery) service 
with unaware participants who, for the most part, have never before used this service. 
Results of the study yield some interesting observations that might be of interests to 
\captcha designers as well as websites using (or considering the use of) \captchas. 

Main contributions of this work are:
\begin{compactitem}
\item A comprehensive quantitative analysis of solving time and how it relates to certain dimensions.
    In particular, this is the first study to obtain multiple solving attempts per person.
    It shows that form-specific checkbox solving time improves with more attempts, with the first attempt 
    being 35\% slower than the 10th, shown in Tables~\ref{tab:st_attempts_checkbox} and 
    \ref{tab:st_attempts_image}. We also show statistically significant changes in checkbox solving 
    time based on the type of service, with password recovery being faster, as shown in
    Tables~\ref{tab:st_service_checkbox}, \ref{tab:st_service_image} and \ref{tab:st_service_total}.
    With respect to educational level\footnote{In the American undergraduate system, "freshmen" 
    are 1st-year students, "sophmore" -- 2nd, "junior" -- 3rd, and "senior" -- 4th.},
    there is a direct trend from freshmen (slowest) to seniors (fastest) at solving \rvii as 
    shown in Tables~\ref{tab:st_sl_checkbox}, \ref{tab:st_sl_image} and \ref{tab:st_sl_total}.
    In terms of participants' major (field of study), there were minor trends with statistical 
    significance of technical (aka STEM) majors solving time being faster than that of non-technical majors,
    as shown in Tables~\ref{tab:st_majors_checkbox}, \ref{tab:st_majors_image} and \ref{tab:st_majors_total}.
 \item An in-depth qualitative analysis of \rvii usability for both checkbox-only and 
    checkbox-and-image combination. Results demonstrate that 40\% of participants found the image 
    version to be annoying (or very annoying), while <10\% found the checkbox version annoying.
    SUS data shows that image results have a mid-score of 58, while checkbox has a score of 78, 
    with 90 being the highest score observed. Based  on the open-ended feedback represented in a 
    {\em word cloud}, participants' most frequent term for the checkbox version
    was {\bf ``easy''} and, for the image version -- {\bf ``annoying''}.
\item A detailed discussion of the cost and security of \rvii (Section~\ref{sec:discussion}).
    Our security analysis shows a blatant vulnerability \cite{the_NOCAPTCHA_problem}, 
    the ease of implementing large-scale automation \cite{Sivakorn2016}, 
    usage of privacy invasive tracking cookies \cite{Sivakorn2016}, 
    and weakness of security premise of fallback (image challenge) ~\cite{attacking_images}. 
    Our cost analysis investigates total human time spent solving \rvii, human labor, 
    network traffic, electricity usage, potential profits and the corresponding environmental impact.
    There have been at least 512  billion \rvii sessions, taking 819 million hours, 
    which translates into at least \$6.1 billion USD in free wages.
    Traffic resulting from \rvii consumed 134 Petabytes of bandwidth, 
    which translates into about 7.5 million kWhs of energy, 
    corresponding to 7.5 million pounds of CO2 pollution.
\end{compactitem}
\noindent {\bf Organization:}  Section~\ref{sec:background} provides some background on  
current \captcha types and System Usability Scale (SUS). Then, Section~\ref{sec:user_study} describes 
the methodology, design, ethics, and implementation of the user study.
Next, Section~\ref{sec:eval} presents the results and their analysis.
Then, Section~\ref{sec:related_work} contextualizes our results against previous user studies.
Next, Section~\ref{sec:discussion} presents the cost and security analysis.
Section~\ref{sec:conclusion} concludes the paper.

	\section{Background} \label{sec:background}
This section overviews the \captcha landscape and System Usability Scale (SUS).
Given familiarity with these topics, it can be skipped with no loss of 
continuity.

\begin{figure}[t]
	\includegraphics[width=\columnwidth]{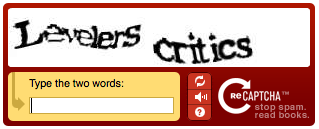}
	\caption{reCAPTCHAv1 distorted text \captcha~\cite{reCAPTCHA}}
	\label{fig:recaptcha_text}
\end{figure}

\begin{figure}[t]
	\includegraphics[width=\columnwidth]{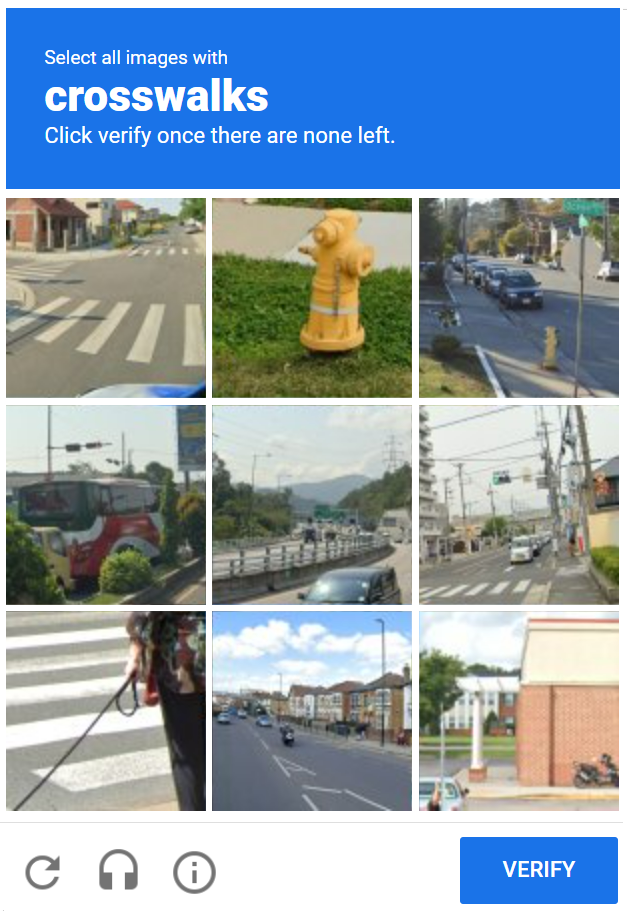}
	\caption{Image Labeling Task \captcha~\cite{reCAPTCHA}}
	\label{fig:recaptcha_image}
\end{figure}
\begin{figure}[t]
	\includegraphics[width=\columnwidth]{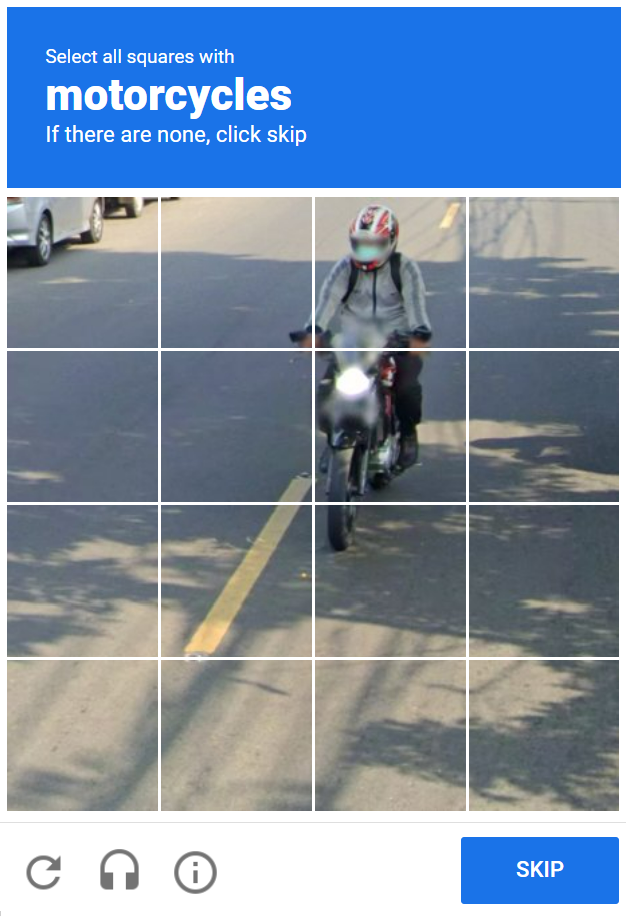}
	\caption{Image Bounding Box Task \captcha~\cite{reCAPTCHA}}
	\label{fig:recaptcha_image_box}
\end{figure}
\begin{figure}[t]
	\includegraphics[width=\columnwidth]{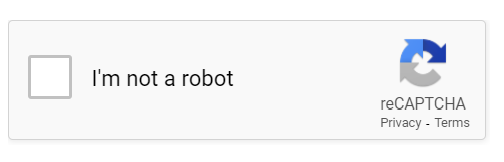}
	\caption{v2 checkbox \captcha~\cite{reCAPTCHAv2}}
	\label{fig:recaptcha_check}
\end{figure}

\subsection{\Captchas}
A recent survey by Guerar et al.~\cite{Guerar} is a comprehensive overview of the current \captcha 
landscape. It proposes a ten-group classification to encompass all current and emerging schemes: 
Text-based, Image-based, Audio-based, Video-based, Game-based, Slider-based, Math-based, 
Behavior-based, Sensor-based, and Liveliness-detection. It also discusses usability, attack resilience, 
privacy, and open challenges for each class. 
Since this paper focuses on behavior and image-based \captchas 
(Which are used in \rvii), we summarize them below. For the rest, we refer to \cite{Guerar}.

\taggedpara{Text-based \captchas} are the earliest type, originally proposed by Von Ahn et al.~\cite{vonAhn}.
They present the user with an image containing a random sequence of visually-distorted alphanumeric characters, 
possibly combined with other visual elements, e.g., lines and dots. The user is required to correctly identify the 
characters and type them into a text field. An example of text-based reCAPTCHA is shown in Figure~\ref{fig:recaptcha_text}.
The idea is that humans should be significantly better than bots at recognizing distorted characters.
However attacks on text-based \captchas have been quite successful and widely studied \cite{Zi20,chen17,tang18,Gao16,li20, Goodfellow}. 
For example, \cite{Zi20} and \cite{li20} achieve over 97\% accuracy on certain text-based \captchas within fractions of a 
second, using machine learning. Furthermore, \cite{Goodfellow} achieves 99.8\% accuracy on reCAPTCHA schemes of the time.
Although research has shown that basic text-based \captchas are no longer effective in distinguishing humans 
from machines, they are still widely used.

\begin{figure}[!ht]
	\includegraphics[width=\columnwidth]{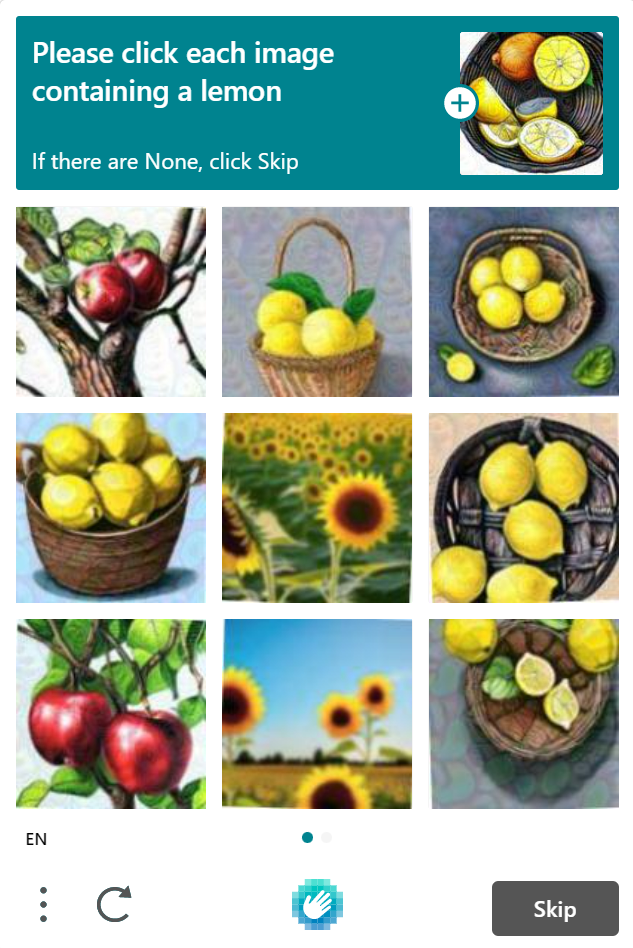}
	\caption{hCAPTCHA~\cite{hCaptcha}}
	\label{fig:hcaptcha}
\end{figure}

\taggedpara{Image-based \captchas} typically require users to perform an image classification task, 
such as selecting images that match the accompanying written description.  Most popular instances
are hCAPTCHA~\cite{hCaptcha} and reCAPTCHA~\cite{reCAPTCHA} version 2 onward. 
Examples are shown in Figure~\ref{fig:recaptcha_image}, Figure~\ref{fig:recaptcha_image_box} and Figure~\ref{fig:hcaptcha}.
The difficulty of these \captchas is associated with that of computer vision-based 
image classification. At the time of the introduction of these \captchas types, corresponding   
problems were not easily solvable by machines. However as computer vision research advanced, 
attacks on image-based \captchas became more successful.
Concrete attacks include \cite{Hossen2020,Hossen2021,Alqahtani2020,Haiqin2019,Lorenzi2012,Sivakorn2016},
some of which report success rates of 85\% for reCAPTCHA and 96\% for hCAPTCHA.

\taggedpara{Behavior-based (or invisible) \captchas} are newer: they either require users to 
click a box (e.g., ``I am not a robot''), or are completely invisible/transparent to the user.
Instead of a visual challenge, they rely on client-side scripts and other opaque techniques 
to collect, in the background, historical behavioral information about the user.
This information is sent to the \captcha provider, which uses various heuristic-based 
techniques to identify bot-like behavior. For instance, Google's popular No-CAPTCHA reCAPTCHA:
\emph{``actively considers a user's entire engagement with the \captcha{} -- before, during, 
and after -- to determine whether that user is a human''}~\cite{reCAPTCHAv2}.
Sivakorn, et al.~\cite{Sivakorn16}, evaluate reCAPTCHA risk analysis system and determine 
that Google tracks cookies, browsing history, and browser environment, e.g., canvas rendering, 
user-agent, screen resolution and mouse. \cite{Sivakorn16} also showed that legitimate cookies can be automatically 
farmed to attack \rvii with 100\% success on a large scale. 

\subsection{System Usability Scale (SUS)}
\begin{figure}[!ht]
    \begin{compactenum} {\footnotesize\bf
        \item[]\hrule 
        \item I think that I would like to use this system frequently.
        \item I found the system unnecessarily complex.
        \item I thought the system was easy to use.
        \item I think that I would need the support of a technical person to be able to use this system.
        \item I found the various functions in this system were well integrated.
        \item I thought there was too much inconsistency in this system.
        \item I would imagine that most people would learn to use this system very quickly.
        \item I found the system very cumbersome to use.
        \item I felt very confident using the system.
        \item I needed to learn a lot of things before I could get going with this system.
        \item[] \hrule
        }
    \end{compactenum}  
    \caption{System Usability Scale (SUS)~\cite{hCaptcha}}
    \label{fig:sus}
\end{figure}

System Usability Scale (SUS), shown in Figure \ref{fig:sus},  is a classical and popular 
survey method designed to assess usability of various systems or products. 
Proposed by Brooke, et al. \cite{brooke1996sus} in 1996, it consists of ten statements:  
five positive and five negative. Each statement is on a 5-point Likert scale 
ranging from \textit{Strongly Disagree (1)} to \textit{Strongly Agree (5)}.  

\noindent
SUS is widely used to measure usability of a wide range of products and systems, from everyday products (such as phones, 
fitness bands, and appliances \cite{kortum2013usability, liang2018usability}) to websites, software, mobile apps and even 
\captchas \cite{kaya2019usability, kortum2013usability, vlachogianni2022perceived, pal2020perceived, klug2017overview, feng2020sencaptcha}. 
SUS is very popular because of its simplicity and conciseness. Participants tend to easily understand and quickly complete the 
SUS questionnaire. The process of calculating scores is also very straightforward:
\begin{compactitem}
    \item For odd-numbered statements, subtract 1 from each response value
    \item For even-numbered statements, subtract each response value from 5
    \item Sum up all response values and multiply the result by 2.5
\end{compactitem}
This yields a SUS score between 0 and 100 for each participant.

To associate a given usability level with individual scores, \cite{bangor2009determining} provides adjective scaling, shown in 
Table \ref{tab: sus adjective scaling}. This scale consists of seven usability levels starting from the worst imaginable 
usability and going up to the best imaginable usability.

\begin{table}[!ht]
\caption{Adjective Ratings of SUS Scores}
\label{tab: sus adjective scaling}
\centering
 \begin{tabularx}{0.7\linewidth}{p{2.5cm} p{2.5cm}}
\toprule  
  Adjective & Mean SUS Score \\
  \midrule
  Worst Imaginable & 12.5 \\ 
  Awful & 20.3 \\
    Poor & 35.7 \\ 
    OK & 50.9 \\ 
    Good & 71.4 \\ 
    Excellent & 85.5 \\ 
  Best Imaginable & 90.9 \\
\bottomrule
\end{tabularx}
\end{table}

\section{The User Study} 
\label{sec:user_study}
Recall that the goals of the user study are to measure solving times, error rates, and user perceptions
of \rvii, the currently prevalent \captcha type. 

\subsection{The Setting \label{setting}}
This study was conducted continuously over the period of roughly 13 months in the 2022-2023 time-frame.
It took place on a campus of a the University of California Irvine, though the scope was limited to one specific
school. The term {\em school} denotes an organizational entity that includes two or more academic departments.
The university contains a number of such schools, e.g., School of Engineering, School of Law, and School of 
Humanities. 

The specific school hosting our study is called SICS: {\em School of Information \& Computer Sciences}
SICS includes several departments, all somehow related to Computer Science. SICS offers a number
of fairly typical undergraduate (BS) and graduate (MS and PhD) programs.

For many years, SICS requires for every person, who for the first time, enrolls in 
any SICS course, to create a SICS-specific user account via the school's web interface.
A typical scenario is that a student who enrolls in at least one SICS course in their 
entire university career, would create a SICS account {\bf only once}.
Consequently, a student who wants to create a SICS account has not previously engaged in 
SICS account creation, meaning that they have no knowledge of the workflow involved,
and no expectations of either seeing or not seeing  \captchas as part of the process.

This motivates the key feature of our user study: introduction (insertion) of \rvii into 
the SICS account management workflow. This actually involves two separate services: 
(1) account creation for new users, and (2) password recovery for users with existing accounts.
This was accomplished with the much-appreciated help and cooperation of the SICS IT Department.

As mentioned earlier, the study ran for about 13 months. This is because we wanted to include
as many  distinct users as possible. Since the yearly academic calendar has multiple terms, we aimed
to catch the beginning of each term (and a week or so prior to it), since this is the time when the 
bulk of new account creation and password recovery activity typically takes place. 

\subsection{Justification}
We now discuss the rationale for the user study setting. Clearly, an ideal and comprehensive
\captcha user study would be as inclusive as possible, comprising a true cross-section of the
world population. Whereas, our study targeted participants are (mostly) university students, 
including undergraduates who range 
from incoming (freshmen) to graduating (seniors), as well as graduate students enrolled
in a variety of programs (MS, MA, MBA, MFA, JD, MD, PhD). The latter are split among so-called 
{\em professional} degree programs, e.g., MBA, JD, MD, and some MS/MA, while others are in regular 
degree programs, e.g., PhD, MFA, and some MS/MA. Such participants are surely are not representative
of the world, or even national, user population. Nonetheless, we conjecture that data stemming
from this admittedly narrow population segment is useful, since it reflects an 
``optimistic'' perception of \captchas. This is because young and tech-savvy users represent the most
agile populations segment and the one most accustomed to dealing with \captchas, due to their heavy Internet use.
Thus, by studying various (not generally positive) impact factors of \captchas, we prefer to err on the 
side of the population that is intuitively the least allergic to \captcha use.

Some reasons for our study setting are fairly obvious. In particular, it would have been very challenging,
if not impossible, to convince any other organization to introduce \captchas into its service workflow, or to allow us
to collect data about their current \captcha use. Alternatively, one could imagine approaching Google and requesting
access to the centralized \rvii service. This would have been ideal since 
it would give us access to a huge number of diverse \rvii users worldwide.
Indeed, we attempted to do this. However Google's legal team denied our request to gain access to large-scale 
data from \rvii. There is very likely a natural counter-incentive for Google (or any other 
\captcha provider) to cooperate with outside researchers in a user study, since doing so might 
reveal certain negative aspects of the service.
Another possibility would have been to create our own brand new service and use \captcha to guard access to it,
thus hoping to attract prospective users of broad demographics.
While theoretically plausible, doing so would be prohibitively time and effort-consuming for
academic researchers. 

Finally, even with our somewhat narrow target demographic of university students, the user study
could have been more latitudinal, i.e., it could span multiple universities in various parts of the world.
This would have yielded more valuable results across political, cultural and linguistic boundaries.
However, this would have been a massive effort requiring careful coordination with, and participation of,
both researchers and IT departments in each university.

\subsection{The Website}
The SICS website used in the study is hosted within the university network.
In order to create a SICS account, a user must first login to the campus VPN 
with their university account. This allows us to claim, with high confidence, that all collected
data stemmed from real human users, who are, for the most part, students (see Section \ref{cleaning} below).

The back-end is a basic PHP server that serves HTML and JavaScript. It is maintained by SICS IT department.
The account creation service includes a form requesting basic student information, e.g., name and student ID.
The password recovery service includes a form requesting existing account information.
In both cases, \rvii was initially hidden and rendered after clicking the submit button.
Basic website workflows for account creation and password recovery are described in Appendix~\ref{sec:workflow}.

All timing events were measured using JavaScript native Date library, which has millisecond 
precision. JavaScript was used to block form submission, such that an initial timing event is 
recorded and a \rvii is rendered simultaneously. Initially, a behavior-based click box 
\captcha (Figure~\ref{fig:recaptcha_check}) is presented. In order to solve it, a user clicks the 
checkbox sending data to Google \rvii site. It either approves the request or presents an 
image-based challenge. Upon \rvii validation, a second timing event is captured and 
the form is submitted.

Solving time is thus comprised of the time interval starting from \captcha rendering until the client browser 
receives a successful validation response from Google \rvii service. (This includes image challenges 
and failed solution attempts.) Upon successful form submission, the IT database stores these two timestamps 
along with the form information.

\subsection{Directory Crawler}
\label{sec: crawler}
Recall that the study involved unwitting participants, i.e., unaware of both existence and purpose of 
the study. In order to subsequently obtain demographic information about each participant, 
we created a JavaScript crawler that automatically searches the university directory using email addresses.
This directory is publicly available from both inside and outside the university network. 
Information gathered by the crawler includes major and college education level (freshman, sophomore, 
junior, senior, or graduate) of each participant.

\subsection{Logistics \& Data Cleaning \label{cleaning}}
In total, the SICS IT department supplied $9,169$ instances of account creation and password recovery 
with \rvii solving time data. The original form data was larger, since it included errors, 
such as incomplete forms and incorrect values.
Each record (form) has the following fields: database ID, date and time, student ID, email address, service, and timing.
Starting with $9,169$ instances, we filtered results using the directory crawler, labeling entries with student 
IDs that were not found and correcting student IDs with minor typos.
A total of $229$ entries were labeled as none for student ID and $295$ student ID typos were corrected.

Successful form submissions have certain constraints, e.g., field formatting.
If a person enters erroneous data that does not fit the constraints, they still have to solve a \rvii before the 
form is submitted. Cases of multiple submissions occurred because of unsuccessful attempts to enter form data.
For some entries, there were small typos, though mixed with temporal evidence they were correctable. 

$28$ records were removed, since each had solving time of $>60$ seconds which adds a high degree of variance. 
We ended up with $9,141$ valid records of which $8,915$ correspond to $3,625$ unique participants.
$226$ entries, labeled as none for student ID, are not included among the unique participants, 
attempts, educational level, and major analysis. Of the $8,915$, $231$ form submissions correspond to 
$52$ unique non-students (i.e., faculty or staff) and are not included in the educational level and 
major analysis. For the purposes of the educational level and major analysis, $3,573$ 
unique students completed $8,631$ \rvii challenges.

\subsection{Post-Study Survey}
After the completion of the study, we randomly selected and contacted, by email, $800$ participants  
in order to solicit feedback on their \rvii experience via a survey (a Google form). In the end, 
a total of $108$ completed the survey. The incentive was an $\$5$ Amazon gift card.
The survey collected answers to SUS questions regarding both checkbox and image \captchas. It also 
collected information about (more detailed) demographics, frequency and nature of internet usage, 
as well as preferences and opinions about checkbox and image \captchas. 

\subsection{Ethical Considerations}
The user study was duly approved by the university's Institutional Review Board (IRB). 
Collection of student email addresses for recruitment and demographic analysis purposes was 
also explicitly approved. Since prospective participants were not pre-informed of their participation in the study, 
two additional documents were filed and approved by the IRB: (1) \emph{``Use of deception/incomplete 
disclosure''} and (2) \emph{``Waiver or Alteration of the Consent''}. 
Study participants who completed the post-study survey were compensated US\$5 for about 5 
minutes of their time. This was also IRB-approved.

No personally identifiable information (PII) was used in the demographics analysis.

After the completion of the study, {\bf all} participants were informed, by email, of 
their participation and the purpose of the study. They were also informed that some basic
demographic information about them that was collected via campus directory lookup.

\section{Results \& Analysis \label{sec:eval}}
This section presents the results of the user study based on the live service experiment.
We consider both quantitative (solving time) and qualitative (SUS, rating, feedback) data 
to provide a comprehensive analysis of \rvii usability.

\subsection{University Demographics}
\begin{figure}
    \includegraphics[width=\columnwidth]{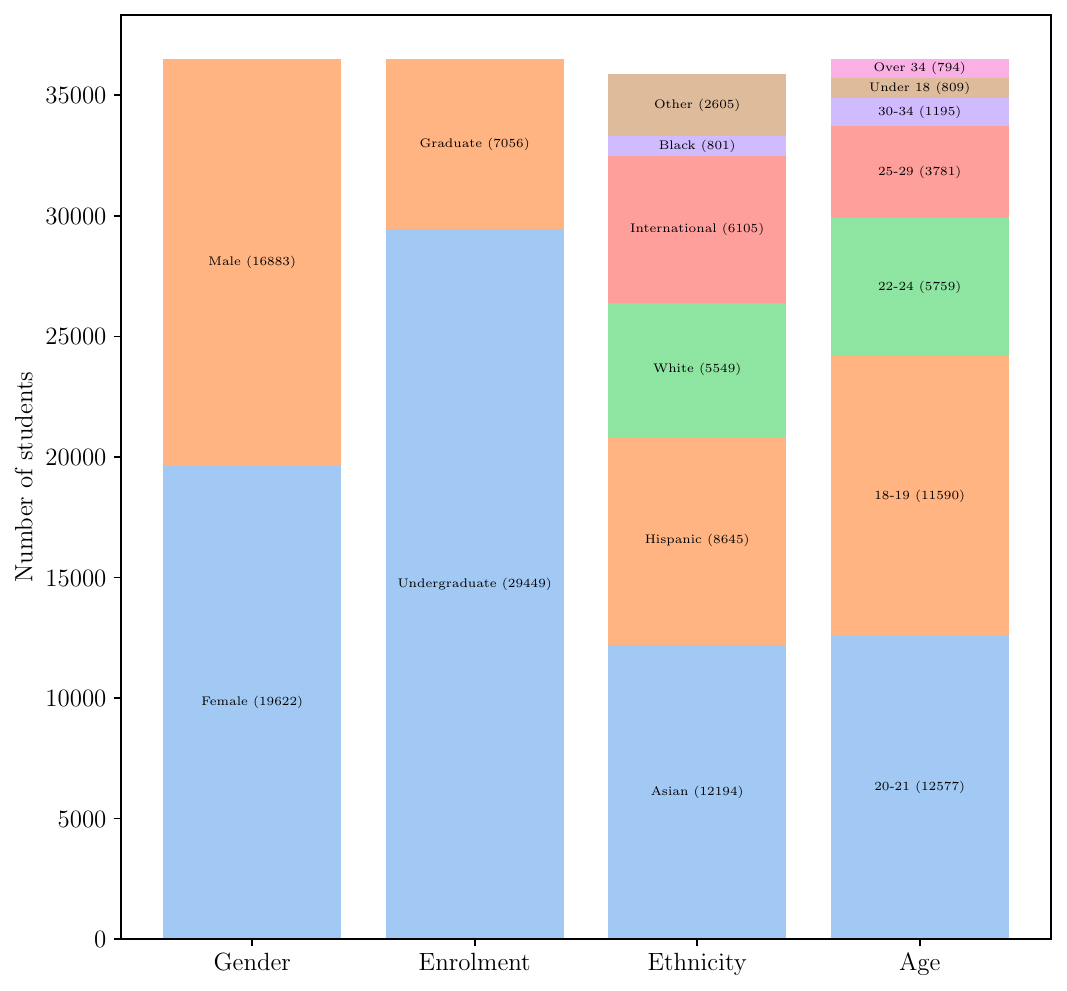}
    \caption{ University Demographics}
    \label{fig: university demographics}
\end{figure}

Student population of the university is large and diverse. 
Figure \ref{fig: university demographics} summarizes its demographics.
We use university demographics, because students from multiple departments who take any 
SICS course create accounts. Thus, demographics about SICS students would not be enough.Moreover, 
the university does not maintain or provide SICS-specific demographics.  

According to recent statistics, the total number of students is $\approx~36,000$ of whom 
54\% are female, 44.6\% are male, plus 1.4\% are non-binary or unstated. In terms of ethnicity, 
the rough breakdown is: 34\% Asian, 24\% Hispanic, 17\% international, 15.44\% White, 2.23\% Black, 
and 7.25\% other ethnic groups. The split between undergraduate and graduate students is 78.10\% to 21.9\%. 

As far as the educational level, freshmen constitute 14\% of the student body, sophomores -- 15\%, 
juniors -- 21\%, and seniors -- 28\%. The rest ($\approx~22\%$) are graduate students. Interestingly, 
the age range of the student population is very wide, ranging from under 18 to over 64. 
Nonetheless, the majority ($82\%$) fall into the $18--24$ age range.

\begin{figure}
    \includegraphics[height=3.5in, width=0.5\textwidth]{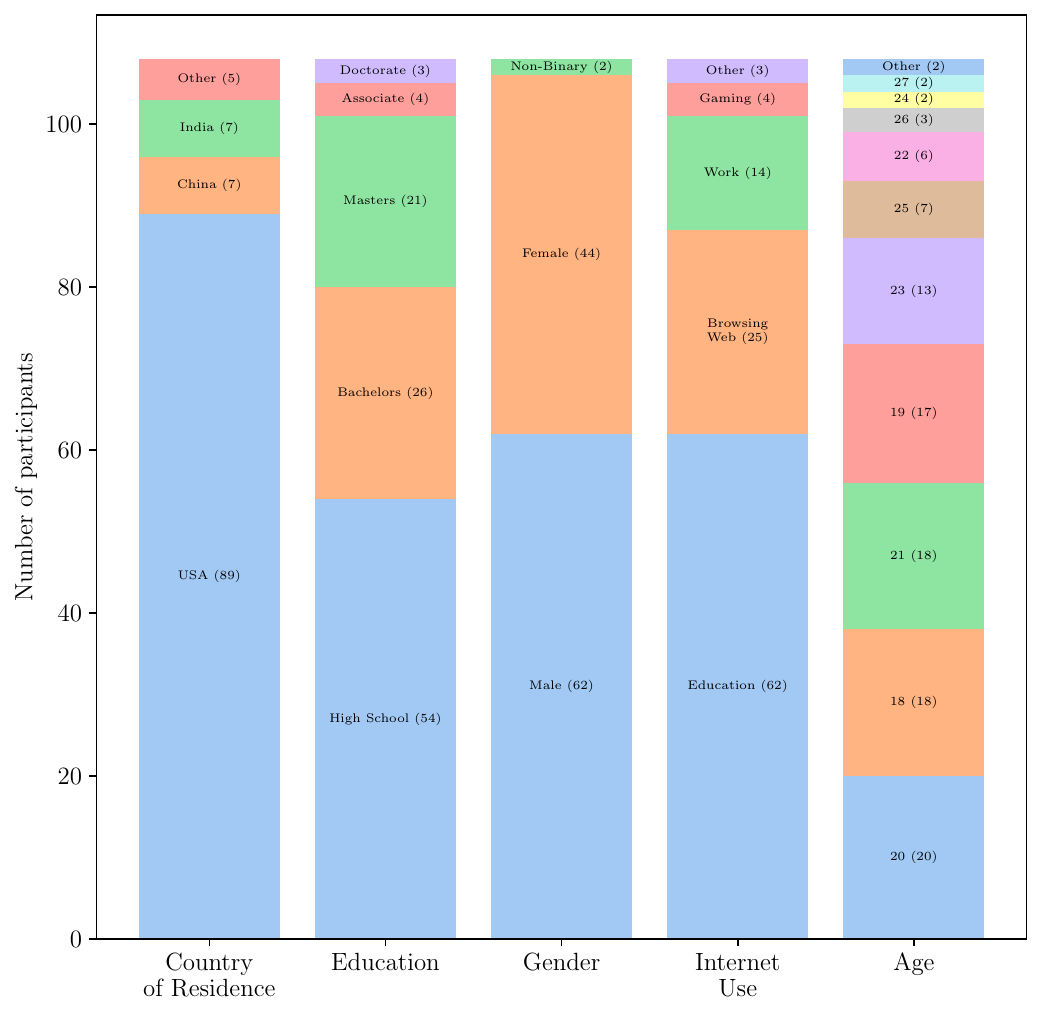}
    \caption{ Demographics of post-study survey participants }
    \label{fig: feedback demographics}
\end{figure}

We also consider demographics of the $108$ participants who engaged in the post-study survey (see 
Figure \ref{fig: feedback demographics}). 
The gender split is $62$ (57.4\%) male, $44$ (40.7\%) female, and $2$ (1.9\%) non-binary. 
The age of participants ranges from $18$ to $30$ with the majority (87.04\%) under 25. 
Participants were also asked about their highest level of education. All participants have at least a 
high school degree. 58 participants (53.7\%) are undergraduates and 50 (46.3\%) are graduate students. 
All participants use the Internet daily and the main purpose of Internet usage for the majority (57.4\%) is 
education. Finally, the country of residence for most (82.4\%) participants is the United States, 
which is directly in line with the 17\% international students from the overall university demographics.

Unfortunately, similarly detailed demographics for participants who solved \rvii as part of the main
live experiment are unknown. However, the demographics of the $108$ who participated in the post-study survey, 
closely resemble those of the overall campus total population in terms of gender, age, and educational level. 
Therefore, it is reasonable to assume that the demographics of all participants are the same, or very similar.

\subsection{\rvii Dashboard Data}
Google provides \rvii analytic data for website operators via a dashboard~\cite{google_admin}.
With it, website operators can generate a key-pair necessary for implementing \rvii on a web page.
Difficulty setting can also be chosen on the dashboard.  We used the ``easy'' setting in all experiments.
The admin console allows for data to be downloaded in CSV format with the following fields per day: 
\begin{quote}
\noindent {\tt\small no CAPTCHAs, Passed CAPTCHAs, Failed CAPTCHAs, Total Sessions, Failed Sessions, 
Average Score, and Average Response Time.}
\end{quote}

\begin{table}[!h]
\centering
\caption{ Google's \rv dashboard data}
\label{tab:dashboard}
\begin{tabularx}{.6\columnwidth}{l | l}
\hline
no CAPTCHAs (checkbox)      &     7629         \\  
Passed CAPTCHAs (Image)  &     1890         \\   
Failed CAPTCHAs (Image)  &     143           \\ 
Total Sessions    &     9538         \\   
Failed Sessions   &     19           \\   
\midrule
Image accuracy    &     92.96\%        \\           
Behavior accuracy &     79.98\%        \\            
\hline
\end{tabularx}
\end{table}

Average score and response time are highly sparse and only appear on days with over $400$ total sessions.
Table~\ref{tab:dashboard} shows a sum for all days when data was collected over the entire study period.
The image accuracy of $93\%$ is computed as: 
$$
(\# passed~ \captchas) \over (\# passed~ \captchas + \# failed~ \captchas)
$$
The behavioral accuracy of $80\%$ is computed as: 
$$ 
( \# of \captchas) \over (total \# sessions)
$$
Notably, there are $9,538$ \captcha sessions reported by the admin console data, 
while we were supplied with $9,169$ sessions, meaning that $369$ form submissions has incomplete data or 
resulted in an error. This is likely due to incomplete sessions, e.g., refreshing before validation, 
or other form submission errors.

\begin{figure}
    \centering
    \includegraphics[height=2.5in, width=\columnwidth]
    {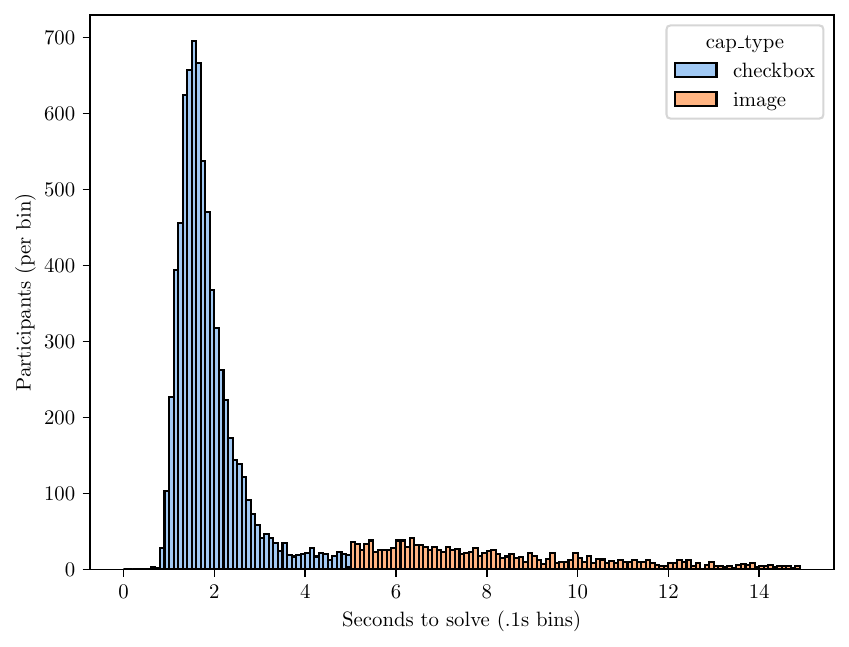}
    \caption{ Timing results in bins of .1 seconds}
    \label{fig:bin_timing}
\end{figure}

\footnotesize
\begin{table}[!ht]
\caption{ Agglomerated solving time for \rv Mode}
\label{tab:st_gen}
\begin{tabularx}{\columnwidth}{l l l l l l l l}
\toprule
mode         &   Count   &      Mean     &    Median    &     Std      &      Var      &     Max    &   Min    \\
\midrule
behavior &   7334 &  1.85 &    1.67 & 0.71 &  0.50 &  4.99 & 0.51 \\
\midrule
image    &   1807 &  10.3 &    8.20 & 6.54 & 42.8  & 59.8  & 4.99 \\
\midrule
total    &   9141 &  3.53 &    1.83 & 4.50 &  20.3 &  59.8 & 0.51 \\
\bottomrule
\end{tabularx}
\end{table}
\normalsize

\subsection{Solving Time}
Solving time for \rvii is measured from the initial display to the successful verification.
Data for solving time is split based on behavioral accuracy of $80\%$ in Table~\ref{tab:dashboard}.
Since all tasks require a checkbox and some also require an image task, we assume that the 
$80\%$ fastest solving times correspond to checkbox interactions.
This split is also noted in the recent work by Searles, et. al~\cite{searlesa_usenix}. 
All timing for image-based results is therefore a combination of check-box and image tasks.

Table~\ref{tab:st_gen} shows the results of $7,334$ behavior and $1,807$ images based on this split.
The mean solving time for behavioral \captchas is $1.85$ seconds, while the image mean solving time 
is $10.3$ seconds. The latter corresponds to a notable 557\% increase. 

\begin{figure}
    \centering
    \includegraphics[height=2.8in,width=\columnwidth]
    {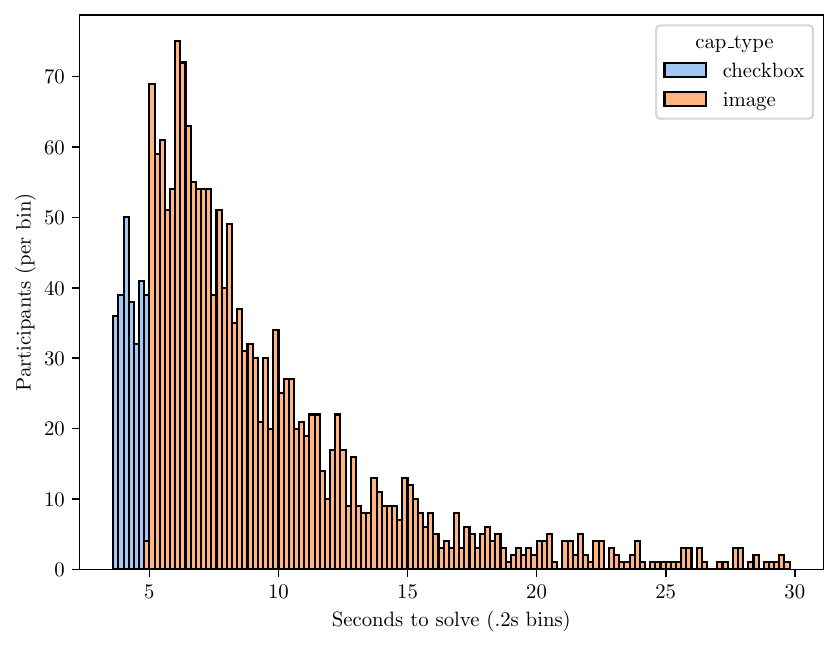}
    \caption{Image timing results in bins of .2 seconds}
    \label{fig:bin_timing_2}
\end{figure}

Looking at Figure~\ref{fig:bin_timing}, there is a sharp drop-off in solving time starting around $2$,
and ending at $5$, seconds:  it hits a low and then goes back up slightly. 
The split point for image and behavior is about $5$ seconds, which matches the drop-off 
point, thus strengthening the accuracy of the split. Figure~\ref{fig:bin_timing_2} shows timing results 
after the image split. Notably, image and checkbox data follow similar patterns of distribution. 

Solving time can also be partitioned into the following dimensions, based on collected data: 
Service, Attempts, Educational Level, and Major. 
This is done separately for image and behavior, across those listed dimensions in 
Tables~\ref{tab:st_service_checkbox}, \ref{tab:st_service_image}, \ref{tab:st_attempts_checkbox}, \ref{tab:st_attempts_image}, 
\ref{tab:st_sl_checkbox}, \ref{tab:st_sl_image}, \ref{tab:st_majors_checkbox} and \ref{tab:st_majors_image}.

\subsubsection{Statistical Testing}
For the sake of statistical validity, we apply a series of standard statistical tests to solving times.
We run all of the following statistical tests on both image and checkbox data separately.
Statistical methods were applied using python's scipy~\cite{scipy} library.
With a null hypothesis that solving times adhere to a normal distribution, we performed the \emph{Shapiro-Wilk normality test}.
For both image and checkbox cases, results showed that we can reject the null hypothesis ($p<0.001$).
With a null hypothesis that the skewness is the same as that of a corresponding normal distribution, we ran the 
timing data with skewtest. For both image and checkbox, results reject the null hypothesis in favor of the 
alternative: the distribution of solving times is skewed ($p<0.001$) to the right. With a null hypothesis that the 
kurtosis is the same as that of a normal distribution, we used the \emph{tailedness test} .
For both image and checkbox, results show the samples were drawn from a population that has a heavy-tailed distribution 
($p<0.001$). We used the \emph{Brown Forsythe test} to compare equality of variance between image and checkbox, 
which shows that they do not exhibit equal variance. We used the \emph{Kruskal-Wallis test} with the Holm-Bonferroni 
method to adjust for family-wise error in order to test the equality of mean between modes, services, attempts, 
majors, and educational level. Significant results are included in 
Figures~\ref{fig:kruskal_a_b}, \ref{fig:kruskal_sl}, and \ref{fig:kruskal_mb}.

\footnotesize
\begin{table}[!ht]
\caption{ Checkbox solving time in seconds for each service}
\label{tab:st_service_checkbox}
\begin{tabularx}{\columnwidth}{l l l l l l l l}
\toprule
 Service   &   Count     &      Mean     &    Median    &     Std      &     Var   &     Max    &   Min     \\
\midrule
 Password Reset &   2654 &  1.67 &    1.51 & 0.65 &  0.42 &  4.99 & 0.51 \\
 Account Creation &   4680 &  1.96 &    1.76 & 0.71 &  0.51 &  4.97 & 0.86 \\
\bottomrule
\end{tabularx}
\end{table}
\normalsize

\footnotesize
\begin{table}[!ht]
\caption{ Image solving time in seconds for each service}
\label{tab:st_service_image}
\begin{tabularx}{\columnwidth}{l l l l l l l l}
\toprule
Service        &   Count     &      Mean     &    Median    &     Std      &     Var   &     Max    &   Min     \\
\midrule
    Password Reset    &    332 & 10.4 &    8.01 & 6.59 & 43.5 & 43.5 & 5.01 \\
    Account Creation    &   1475 & 10.3 &    8.23 & 6.53 & 42.7 & 59.8 & 4.99 \\
\bottomrule
\end{tabularx}
\end{table}
\normalsize

\footnotesize
\begin{table}[!ht]
\caption{ Total solving time in seconds for each service}
\label{tab:st_service_total}
\begin{tabularx}{\columnwidth}{l l l l l l l l}
\toprule
Service        &   Count     &      Mean     &    Median    &     Std      &     Var   &     Max    &   Min     \\
\midrule
    Password Reset    &   2986 &  2.63 &    1.58 & 3.56 & 12.7 & 43.5 & 0.51 \\
    Account Creation  &   6155 &  3.97 &    2.00 & 4.84 & 23.4 & 59.8 & 0.86 \\
\bottomrule
\end{tabularx}
\end{table}
\normalsize

\subsubsection{Services}
As mentioned earlier, the website had two services that invoked \rvii: password recovery and account creation.
Tables~\ref{tab:st_service_checkbox}, \ref{tab:st_service_image}, and \ref{tab:st_service_total} show results from 
these two \captcha interactions.
There were $6,155$ account creation, and $2,986$ password recovery, form submissions. 
Notably, for behavioral results, the Kruskal-Wallace test shows statistically significant differences between account 
creation and password recovery with a $p=1.1e^{-115}$.
Students who interacted with the account creation service solved behavioral \captchas $17\%$ slower than those who 
interacted with the password recovery service.
Additionally, $50\%$ more time was spent solving \rvii during account creation than during password recovery.
Total results are also statistically significant with $p=6.7e^{-162}$.
However, since 90\% of students who interacted with the latter have already interacted with the account creation service,
these results may be conflated by multiple prior attempts.
For the image case, the Kruskal-Wallace Test yielded no statistically significant results.

\footnotesize
\begin{table}[!htbp]
\caption{ Solving time for number of checkbox attempts}
\label{tab:st_attempts_checkbox}
\begin{tabularx}{\columnwidth}{l l l l l l l l}
\toprule
Attempt     &   Count     &      Mean     &    Median     &     Std        &     Var    &     Max    &   Min     \\
\midrule
    1  &   2888 &  2.02 &    1.80 & 0.73 &  0.54 &  4.97 & 0.94 \\
    2  &   1293 &  1.84 &    1.67 & 0.65 &  0.42 &  4.97 & 0.62 \\
    3  &    751 &  1.80 &    1.63 & 0.66 &  0.44 &  4.95 & 0.80 \\
    4  &    513 &  1.73 &    1.55 & 0.63 &  0.40 &  4.89 & 0.78 \\
    5  &    371 &  1.73 &    1.57 & 0.70 &  0.49 &  4.92 & 0.89 \\
    6  &    272 &  1.61 &    1.47 & 0.58 &  0.34 &  4.57 & 0.84 \\
    7  &    212 &  1.67 &    1.52 & 0.65 &  0.43 &  4.90 & 0.64 \\
    8  &    167 &  1.66 &    1.52 & 0.65 &  0.43 &  4.65 & 0.84 \\
    9  &    127 &  1.60 &    1.48 & 0.57 &  0.33 &  4.09 & 0.88 \\
    10 &    112 &  1.56 &    1.44 & 0.63 &  0.39 &  4.97 & 0.85 \\
    11 &     94 &  1.63 &    1.41 & 0.76 &  0.57 &  4.90 & 0.88 \\
    12 &     67 &  1.61 &    1.46 & 0.68 &  0.46 &  4.47 & 0.51 \\
    13 &     52 &  1.58 &    1.37 & 0.70 &  0.49 &  4.49 & 0.96 \\
    14 &     37 &  1.53 &    1.45 & 0.63 &  0.40 &  4.62 & 0.92 \\
    15 &     28 &  1.51 &    1.41 & 0.56 &  0.31 &  3.88 & 0.88 \\
\bottomrule
\end{tabularx}
\end{table}

\footnotesize
\begin{table}[!ht]
\caption{ Solving time for number of image attempts}
\label{tab:st_attempts_image}
\begin{tabularx}{\columnwidth}{l l l l l l l l}
\toprule 
Attempt   &   Count   &      Mean     &    Median     &     Std         &    Var      &     Max     &    Min     \\
\midrule
 1  &   1264 &  10.5 &    8.36 & 6.60 & 43.5 & 58.9 & 4.99 \\
 2  &    260 &  10.9 &    8.16 & 7.47 & 55.8 & 55.5 & 5.00 \\
 3  &     93 &  9.30 &    8.16 & 4.09 & 16.7 & 29.2 & 5.00 \\
 4  &     45 &  10.0 &    7.77 & 8.41 & 70.7 & 59.8 & 5.21 \\
 5  &     25 &  8.76 &    7.48 & 4.56 & 20.8 & 26.4 & 5.12 \\
 6  &     15 &  7.26 &    6.06 & 2.33 & 5.44 & 12.3 & 5.18 \\

\bottomrule
\end{tabularx}
\end{table}
\normalsize

\subsubsection{Attempts}
\begin{figure}
    \centering
    \includegraphics[width=\columnwidth]
    {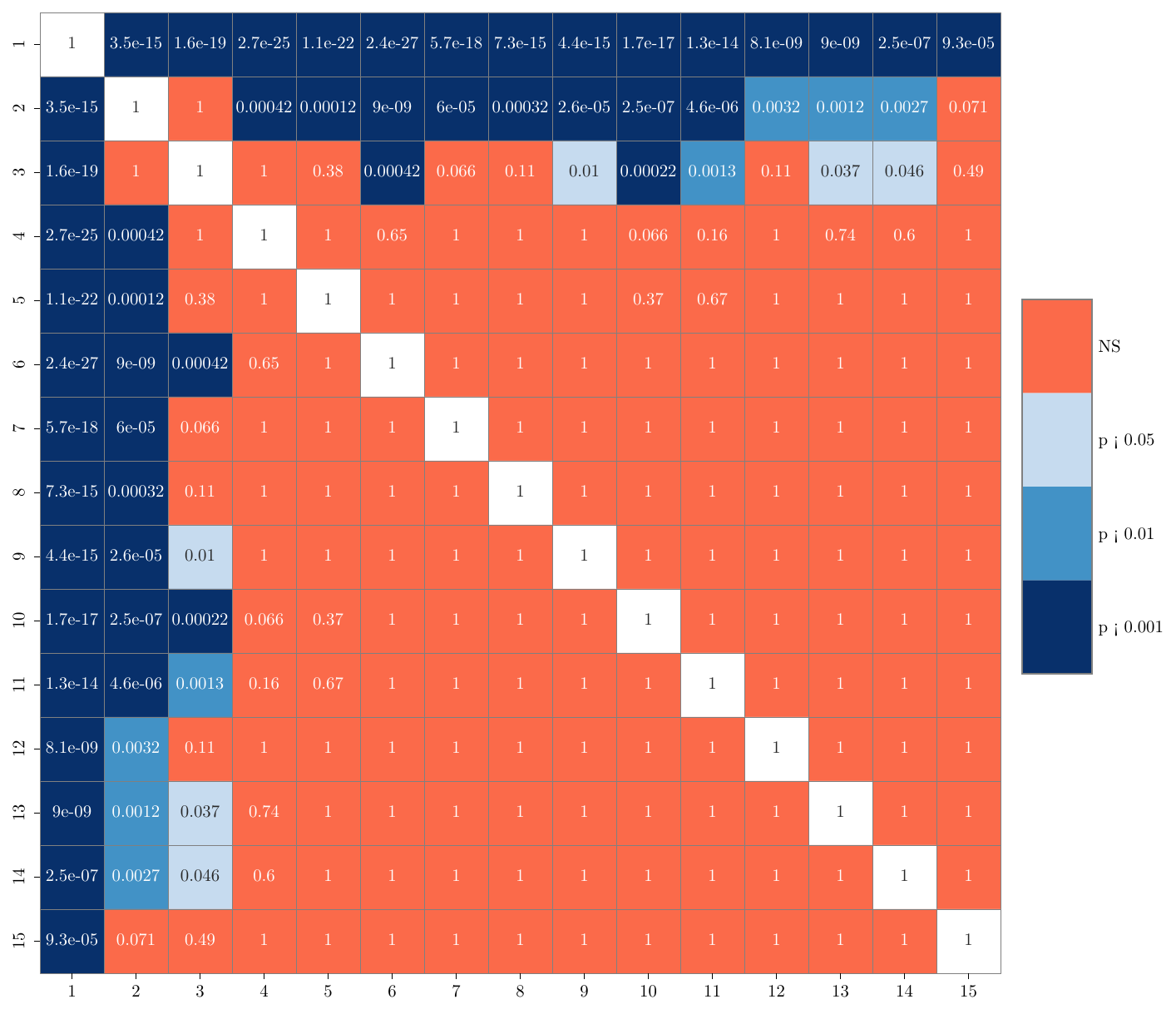}
    \caption{Kruskal-Wallace results for checkbox attempts}     
    \label{fig:kruskal_a_b}
\end{figure}

Interestingly, some participants submitted forms multiple times. For behavior-based 
challenges, the average number of attempts was $3.52$, and $1.73$ for image-based ones.
Tables~\ref{tab:st_attempts_checkbox} and \ref{tab:st_attempts_image} show  timing results over multiple attempts.
The highest number of attempts was $37$ for behavior-based challenges and $20$ for image-based ones. 
Behavioral results from the Kruskal-Wallis test in Figure~\ref{fig:kruskal_a_b} show that there is a 
statistically significant difference between the first and subsequent attempts $(p<.001)$. 
While for the second attempt there is a statistically significant difference $(p<.001)$ between all 
other attempts except the third.  In general, this data shows that checkbox solving time 
decreases with more attempts, meaning that humans improve at solving checkbox challenges.

We observe an interesting behavioral phenomena whereby participants react faster when they know what to expect.
However, average image results show a slight increase on the second attempt, while subsequent attempts decrease.
This may be attributed to \rvii presenting a more difficult challenge on the second attempt.
Image results from the Kruskal-Wallis test show no statistically significant differences between image attempts. 
This is likely due to the drop-off in the number of participants who solved multiple image challenges.

\footnotesize
\begin{table}[!ht]
\caption{Checkbox solving time for educational levels}
\label{tab:st_sl_checkbox}
\begin{tabularx}{\columnwidth}{l l l l l l l l l}
\toprule
  Level      &   Count    &      Mean     &    Median    &     Std      &     Var   &     Max    &   Min &  Total \%   \\
\midrule
Freshmen  &    479 &  1.89 &    1.68 & 0.77 &  0.59 &  4.96 & 0.95 & 62.1\% \\
Sophomore &   1198 &  1.93 &    1.73 & 0.73 &  0.54 &  4.99 & 0.91 & 71.3\% \\
Junior    &   1912 &  1.84 &    1.66 & 0.69 &  0.48 &  4.96 & 0.51 & 85.2\% \\
Senior    &   2399 &  1.79 &    1.63 & 0.68 &  0.46 &  4.95 & 0.64 & 87.4\% \\
Graduate  &    932 &  1.95 &    1.76 & 0.70 &  0.49 &  4.96 & 0.91 & 78.6\% \\
\bottomrule
\end{tabularx}
\end{table}
\normalsize        

\footnotesize
\begin{table}[!ht]
\caption{Image solving time for different educational levels}
\label{tab:st_sl_image}
\begin{tabularx}{\columnwidth}{l l l l l l l l l}
\toprule
 Level      &   Count     &      Mean     &    Median    &     Std      &     Var   &     Max    &   Min  &  Total \% \\
\midrule
Freshmen  &    294 & 10.5 &    8.42 & 6.24 & 38.9 & 56.2 & 5.01 & 37.9\% \\
Sophomore &    483 & 10.3 &    8.08 & 7.28 & 52.9 & 59.8 & 5.00 & 28.7\% \\
Junior    &    334 & 10.3 &    8.18 & 6.00 & 36.0 & 45.4 & 5.00 & 14.8\% \\
Senior    &    346 & 10.2 &    8.05 & 6.20 & 38.5 & 43.9 & 4.99 & 12.6\% \\
Graduate  &    254 & 10.7 &    8.49 & 6.85 & 46.9 & 50.0 & 5.01 & 21.4\% \\
\bottomrule
\end{tabularx}
\end{table}
\normalsize

\footnotesize
\begin{table}[!ht]
\caption{Total solving time for different educational levels}
\label{tab:st_sl_total}
\begin{tabularx}{\columnwidth}{l l l l l l l l}
\toprule
 Level      &   Count     &      Mean     &    Median    &     Std      &     Var   &     Max    &   Min     \\
\midrule
Freshmen  &    773 &  5.15 &    2.33 & 5.69 & 32.4 & 56.2 & 0.95 \\
Sophomore &   1681 &  4.33 &    2.05 & 5.47 & 29.9 & 59.8 & 0.91 \\
Junior    &   2246 &  3.09 &    1.77 & 3.84 & 14.7 & 45.4 & 0.51 \\
Senior    &   2745 &  2.85 &    1.71 & 3.62 & 13.1 & 43.9 & 0.64 \\
Graduate  &   1186 &  3.82 &    1.97 & 4.83 & 23.3 & 50.0 & 0.91 \\
\bottomrule
\end{tabularx}
\end{table}
\normalsize

\footnotesize
\begin{table}[!ht]
\caption{ Checkbox solving time for various majors}
\label{tab:st_majors_checkbox}
\begin{tabularx}{\columnwidth}{l l l l l l l l}
\toprule
Major      &   Count   &     Mean   &     Median   &     Std       &     Var       &     Max      &   Min     \\
\midrule
CmptSci  &   2658 &  1.80 &    1.63 &  0.66 &   0.44 &  4.99 & 0.62 \\
CSE     &    754 &  1.83 &    1.63 &  0.77 &   0.59 &  4.97 & 0.64 \\
SW Engr   &    635 &  1.79 &    1.59 &  0.70 &   0.50 &  4.92 & 0.51 \\
Undclrd   &    613 &  1.91 &    1.70 &  0.75 &   0.56 &  4.97 & 0.95 \\
MCS      &    326 &  2.01 &    1.85 &  0.68 &   0.47 &  4.96 & 0.91 \\
DataSci  &    273 &  1.98 &    1.74 &  0.80 &   0.65 &  4.96 & 1.03 \\
IN4MATX   &    225 &  1.88 &    1.70 &  0.66 &   0.44 &  4.96 & 1.01 \\
BIM     &    171 &  1.96 &    1.73 &  0.81 &   0.66 &  4.94 & 0.89 \\
GameDes   &    139 &  1.80 &    1.60 &  0.74 &   0.55 &  4.79 & 0.77 \\
Math     &    121 &  1.84 &    1.72 &  0.59 &   0.35 &  3.93 & 1.00 \\
MofData   &    102 &  1.90 &    1.70 &  0.72 &   0.51 &  4.60 & 1.03 \\
EngrCpE   &     82 &  1.90 &    1.74 &  0.66 &   0.44 &  4.36 & 0.98 \\
PSW ENG   &     79 &  2.03 &    1.81 &  0.72 &   0.52 &  4.69 & 0.91 \\
Bus Adm  &     77 &  1.92 &    1.77 &  0.78 &   0.61 &  4.78 & 0.88 \\
CSGames  &     77 &  1.73 &    1.55 &  0.66 &   0.44 &  4.76 & 0.88 \\
BusEcon  &     62 &  1.98 &    1.77 &  0.66 &   0.44 &  4.95 & 0.95 \\
Bio Sci  &     60 &  1.89 &    1.69 &  0.74 &   0.55 &  4.89 & 0.99 \\
Stats    &     47 &  1.73 &    1.56 &  0.61 &   0.37 &  4.59 & 1.11 \\
Cog Sci  &     37 &  1.97 &    1.94 &  0.56 &   0.31 &  3.39 & 0.97 \\
Net Sys   &     34 &  2.13 &    1.96 &  0.69 &   0.47 &  3.92 & 1.33 \\
Psych    &     29 &  1.70 &    1.59 &  0.44 &   0.19 &  2.62 & 1.05 \\
Engr ME   &     26 &  2.06 &    1.85 &  0.54 &   0.29 &  3.44 & 1.39 \\
\bottomrule
\end{tabularx}
\end{table}
\normalsize

\subsubsection{Educational Level}
\begin{figure}[!ht]
    \centering
    \includegraphics[height=3.2in,width=\columnwidth]
    {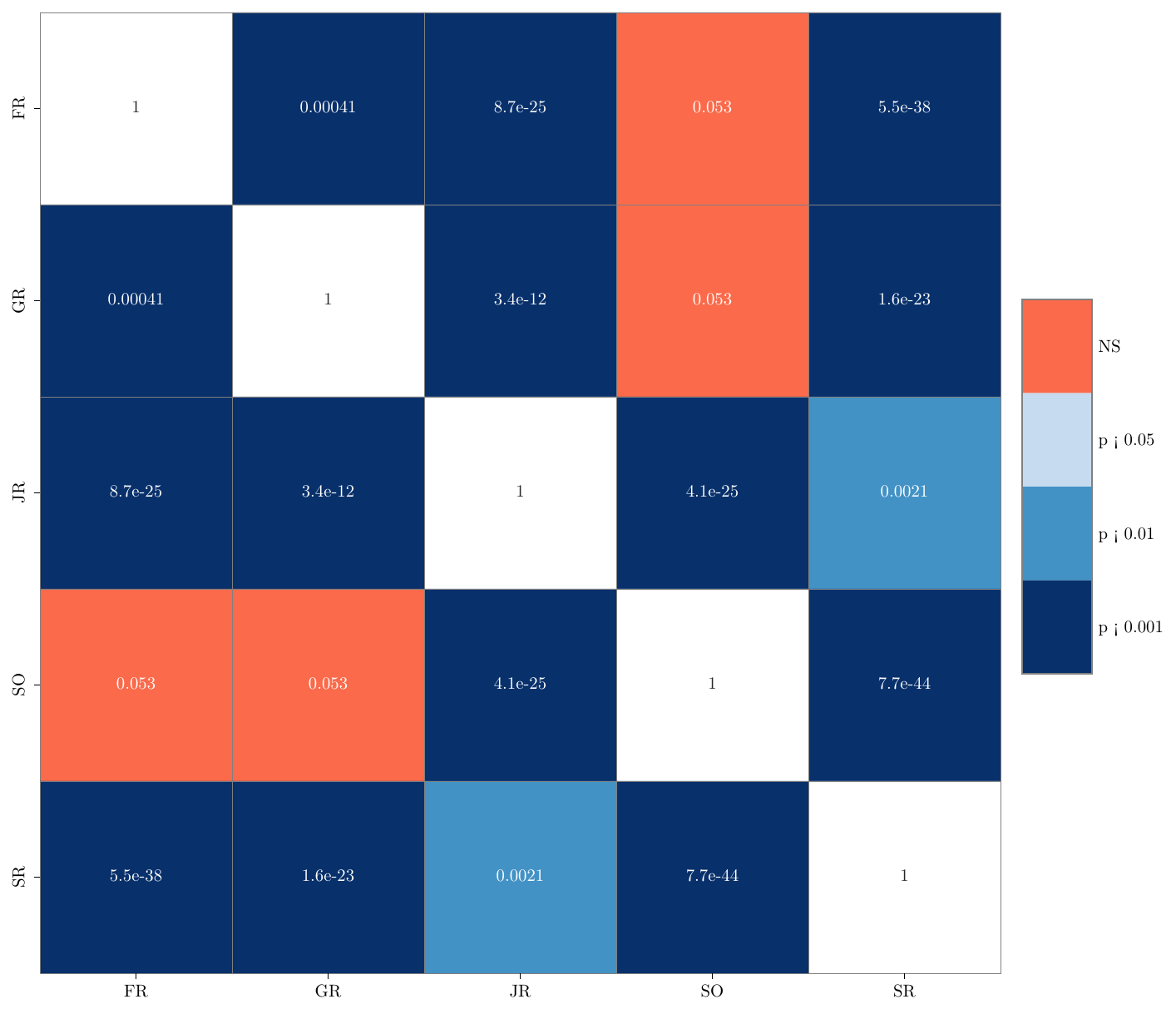}
    \caption{Kruskal-Wallace results for total and educational level }
    \label{fig:kruskal_sl}
\end{figure}

Educational level was obtained via the website crawler, as described in Section~\ref{sec: crawler}.
Tables~\ref{tab:st_sl_checkbox}, \ref{tab:st_sl_image}, and \ref{tab:st_sl_total} present data for different educational levels.
For the split checkbox and image data there are only very minor differences between solving times based on the educational level.
In terms of statistical significance, Figure~\ref{fig:kruskal_sl} shows statistically significant differences in total 
solving time for all educational levels. In terms of total time, freshmen are the slowest -- 
80\% slower than seniors. There is a direct trend from freshman to seniors showing a reduction in solving time.
Similarly, there is a trend of the total ratio of image to checkbox challenges.

\footnotesize
\begin{table}[!ht]
\caption{ Image solving time for various majors}
\label{tab:st_majors_image}
\begin{tabularx}{\columnwidth}{l l l l l l l l}
\toprule
Major      &   Count   &     Mean     &    Median  &     Std        &     Var   &     Max     &   Min     \\
\midrule
\small
CmptSci  &    527 & 10.2 &    8.13 &  6.12 &  37.4 & 44.5 & 4.99 \\
Undclrd  &    239 & 11.0 &    8.15 &  8.25 &  68.0 & 59.8 & 5.02 \\
CSE      &    196 &  9.94 &    8.26 &  5.70 &  32.5 & 42.0 & 5.03 \\
SW Engr  &    161 &  9.64 &    7.63 &  5.51 &  30.3 & 45.4 & 5.04 \\
DataSci  &     90 & 10.2 &    8.51 &  5.84 &  34.1 & 41.2 & 5.01 \\
MCS      &     78 & 10.5 &    8.65 &  6.01 &  36.2 & 38.1 & 5.05 \\
IN4MATX  &     62 & 12.4 &    8.37 &  9.82 &  96.4 & 50.9 & 5.14 \\
BIM      &     55 &  9.47 &    8.35 &  4.25 &  18.0 & 25.5 & 5.02 \\
GameDes  &     47 & 12.0 &    8.58 & 10.8 & 117 & 56.2 & 5.16 \\
MofData  &     29 &  9.71 &    8.99 &  4.48 &  20.1 & 25.6 & 5.39 \\
Math     &     28 & 10.6 &    8.98 &  5.08 &  25.8 & 28.9 & 5.34 \\
EngrCpE  &     24 & 10.9 &    9.11 &  4.27 &  18.2 & 20.1 & 5.42 \\
PSW ENG  &     18 &  8.98 &    7.58 &  4.30 &  18.5 & 21.3 & 5.01 \\
Stats    &     18 &  8.84 &    6.89 &  4.60 &  21.1 & 23.9 & 5.73 \\
BusEcon  &     16 &  9.74 &    9.39 &  4.45 &  19.8 & 21.3 & 5.11 \\
Bio Sci  &     15 & 12.0 &   11.2 &  5.70 &  32.5 & 23.0 & 5.72 \\
Bus Adm  &     12 &  8.78 &    8.87 &  1.87 &   3.51 & 13.1 & 5.56 \\
CSGames  &      8 & 10.7 &   10.1 &  4.04 &  16.4 & 16.1 & 5.90 \\
Engr ME  &      8 &  9.26 &    7.85 &  4.20 &  17.6 & 18.4 & 5.38 \\
Cog Sci  &      7 &  8.15 &    6.66 &  4.17 &  17.4 & 16.8 & 5.00 \\
Net Sys  &      5 & 21.0 &    9.47 & 18.5 & 343 & 50.0 & 7.16 \\
Pysch    &      3 &  6.41 &    6.29 &  1.18 &   1.40 &  7.65 & 5.30 \\
\bottomrule
\end{tabularx}
\end{table}
\normalsize

\footnotesize
\begin{table}[!ht]
\caption{ Total solving time for various majors}
\label{tab:st_majors_total}
\begin{tabularx}{\columnwidth}{l l l l l l l l}
\toprule
Major      &   Count   &     Mean     &    Median  &     Std        &     Var   &     Max     &   Min     \\
\midrule
\small
CmptSci   &   3185 &  3.19 &    1.75 & 4.05 &  16.4 & 44.5 & 0.62 \\
CSE       &    950 &  3.51 &    1.81 & 4.23 &  17.9 & 42.0 & 0.64 \\
Undclrd   &    850 &  4.47 &    2.03 & 6.02 &  36.2 & 59.8 & 0.95 \\
SW Engr   &    796 &  3.38 &    1.75 & 4.06 &  16.5 & 45.4 & 0.51 \\
MCS       &    404 &  3.65 &    2.08 & 4.32 &  18.7 & 38.1 & 0.91 \\
DataSci   &    362 &  3.98 &    2.02 & 4.55 &  20.7 & 41.2 & 1.03 \\
IN4MATX   &    287 &  4.14 &    1.89 & 6.29 &  39.5 & 50.9 & 1.01 \\
BIM       &    226 &  3.79 &    1.97 & 3.91 &  15.3 & 25.5 & 0.89 \\
GameDes   &    186 &  4.38 &    1.86 & 7.03 &  49.4 & 56.2 & 0.77 \\
Math      &    147 &  3.50 &    1.89 & 4.11 &  16.9 & 28.9 & 1.00 \\
MofData   &    131 &  3.63 &    1.94 & 3.92 &  15.3 & 25.6 & 1.03 \\
EngrCpE   &    106 &  3.93 &    1.98 & 4.31 &  18.6 & 20.1 & 0.98 \\
PSW ENG   &     97 &  3.32 &    1.93 & 3.33 &  11.1 & 21.3 & 0.91 \\
Bus Adm   &     89 &  2.85 &    1.83 & 2.55 &  6.52 & 13.1 & 0.88 \\
CSGames   &     85 &  2.43 &    1.61 & 2.60 &  6.77 & 18.4 & 0.88 \\
BusEcon   &     78 &  3.57 &    2.06 & 3.76 &  14.1 & 21.3 & 0.95 \\
Bio Sci   &     75 &  3.90 &    1.87 & 4.80 &  23.0 & 23.0 & 0.99 \\
Stats     &     65 &  3.70 &    1.68 & 4.02 &  16.2 & 23.9 & 1.11 \\
Cog Sci   &     44 &  2.96 &    2.02 & 2.81 &  7.90 & 16.8 & 0.97 \\
Net Sys   &     39 &  4.55 &    2.07 & 8.81 &  77.6 & 50.0 & 1.33 \\
Engr ME   &     34 &  4.08 &    2.16 & 4.18 &  17.5 & 16.1 & 1.39 \\
Psych     &     32 &  2.14 &    1.65 & 1.49 &  2.21 & 7.65 & 1.05 \\
\bottomrule
\end{tabularx}
\end{table}
\normalsize

\subsubsection{Majors}
\begin{figure}
    \centering
    \includegraphics[width=\columnwidth]
    {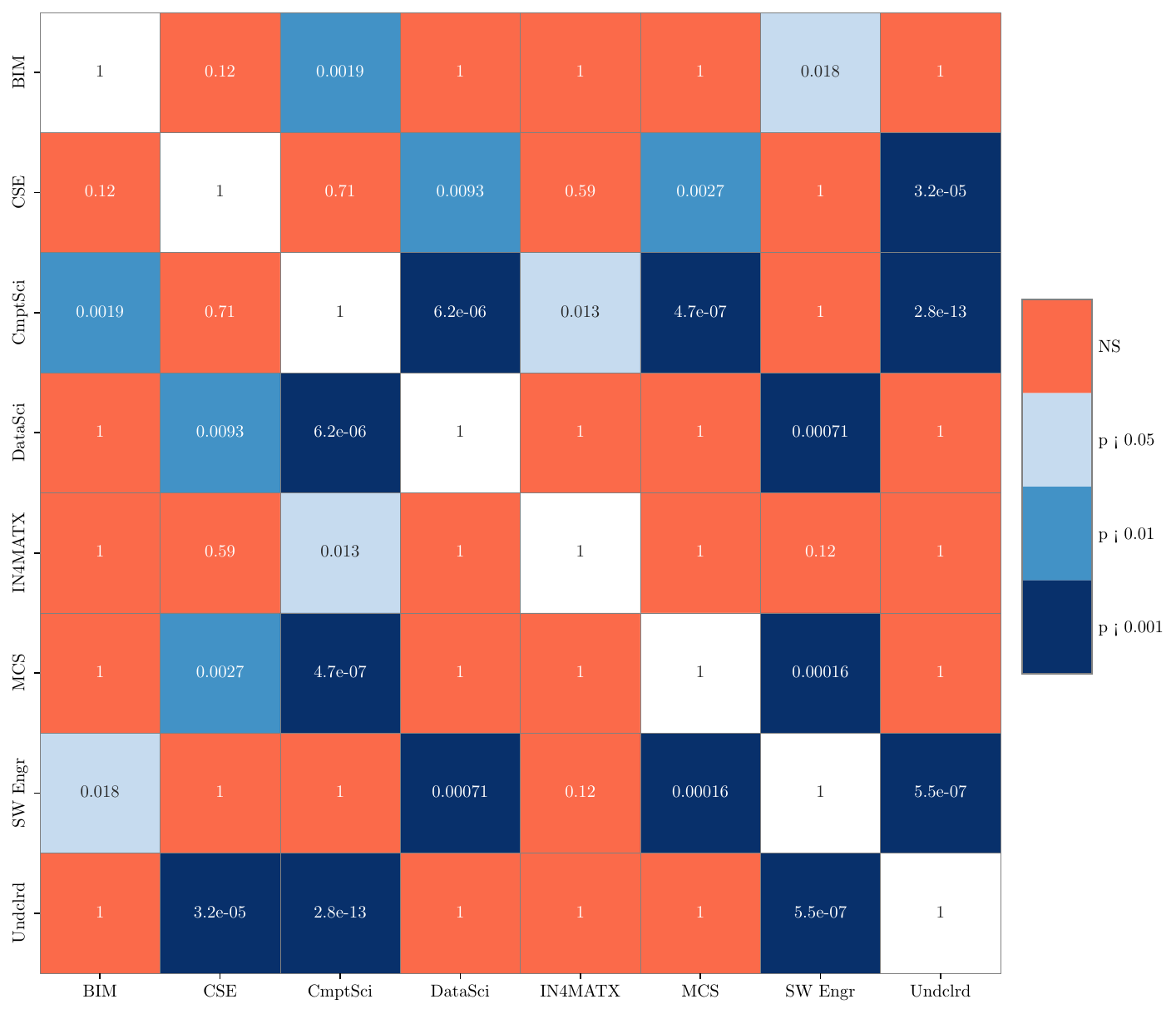}
    \caption{Kruskal-Wallace results for total and major}
    \label{fig:kruskal_mb}
\end{figure}

Majors of the study participants (i.e., disciplines they study) were obtained through 
the website crawler, as described in Section~\ref{sec: crawler}.
Tables~\ref{tab:st_majors_checkbox}, \ref{tab:st_majors_total}, and \ref{tab:st_majors_image} present solving times 
for participants with various majors. Although there are 62 majors in total, 
Tables~\ref{tab:st_majors_checkbox}, \ref{tab:st_majors_total}, and \ref{tab:st_majors_image} only show 22 majors. This is because
each of the remaining 40 majors had $<20$ \rvii sessions.
As the Kruskall-Wallace test in Figure~\ref{fig:kruskal_mb} shows, only 8 majors had 
statistically significant differences in terms of checkbox solving behavior.
Among these, Computer Science had the lowest, and Informatics -- the highest, total average solving time.

\subsection{Survey Results}
We now discuss the study results pertaining to usability, preferences, and opinions about \rvii.
An interactive version of the google form we used is available at \cite{survey_form}.
$800$ randomly selected study participants were contacted by email, with the goal of obtaining at least $100$ 
respondents. In the end, a total of $108$ completed the survey. 
Two solving scenarios are considered:
\begin{description}
\item[Checkbox only] Only the checkbox challenge: after clicking 
the checkbox, no image challenge was served. This applies to $42$ participants.
\item[Checkbox+image] Both checkbox and image challenges: after 
clicking the checkbox, an image challenge was served. This applies to $66$ participants.
\end{description}
\subsubsection{System Usability Scale (SUS) Score Analysis}
Table \ref{tab: reCAPTCHA sus score} reports the SUS score for both scenarios. 
Results from individual SUS statements are not analyzed, since they do not provide 
meaningful information \cite{bangor2009determining, brooke1996sus}.

\begin{table}[t!]
\caption{SUS Scores for \rvii}
\label{tab: reCAPTCHA sus score}
\centering
 \begin{tabularx}{\linewidth}{lll}
\toprule  
\textbf{Solving Scenario} & \textbf{\rv Type} & \textbf{SUS Score} \\
\midrule
Checkbox only & Checkbox & 78.51 \\ 
Checkbox+image &  Checkbox & 76.21 \\
Checkbox+image &  Image & 58.90 \\
\bottomrule
\end{tabularx}
\vspace{1.5em}
\end{table}

SUS checkbox scores are: $78.51$ for checkbox only, and $76.21$ for checkbox+image. 
Referring to Table \ref{tab: sus adjective scaling}, the usability level for checkbox 
in both scenarios is ``Good''. We thus conclude that for checkbox, the SUS score and the 
usability level do not vary depending on the solving scenario, i.e., whether or not 
an image challenge is served afterwards.  On the other hand, the SUS score of image 
is $58.90$ and the usability level is ``OK''.  This difference is likely 
influenced by the difficulty of the task, since clicking a checkbox is surely 
much simpler than classifying an image. We observed that solving image 
challenges takes $557\%$ longer than checkbox.

\subsubsection{Preference Analysis}
\begin{figure}
    \includegraphics[width=0.5\textwidth]{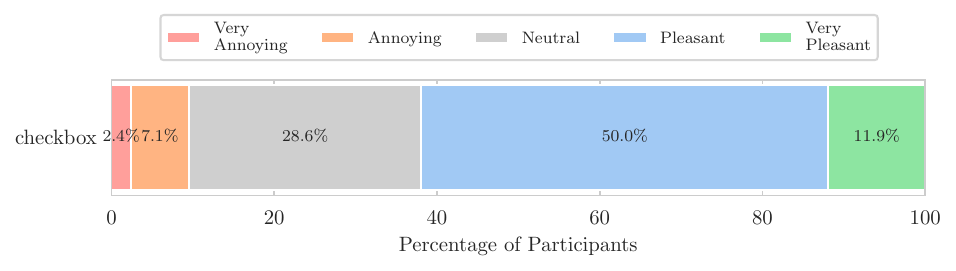}
    \caption{Preference score for checkbox only scenario} 
    \label{fig: checkbox only preference}
\end{figure}

\begin{figure}
    \includegraphics[width=0.5\textwidth]{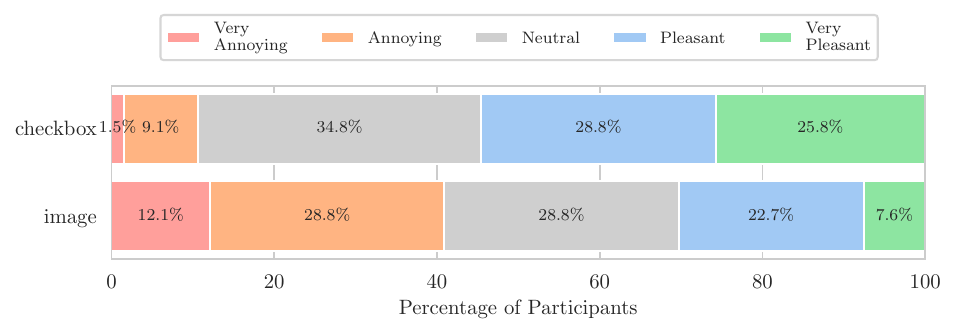}
    \caption{Preference score for checkbox+image scenario} 
    \label{fig: checkbox+image preference}
\end{figure}

Besides SUS questions, the post-study questionnaire asked about participants' preferences 
regarding checkbox and image versions. Specifically, they were asked to provide 
opinions using a custom scale. Figure \ref{fig: checkbox only preference} and Figure 
\ref{fig: checkbox+image preference} show the preferences in both scenarios.  

Majority of participants in both scenarios (61.9\% and 54.6\%, respectively) 
find checkbox either ``pleasant'' or ``very pleasant''. Whereas, only a minority (30.3\%) 
find image ``pleasant'' or ``very pleasant''. A significantly larger percentage of 
participants think that image is ``annoying'' or ``very annoying''.
Similar to SUS scores, the preference for checkbox does not change when it is followed by an image. 

We also compute quantitative scores in order to rate \rvii based on participants' preferences.
For that, the preference scale is converted into a five-point Likert scale with ``very annoying''
corresponding to 1 and ``very pleasant'' corresponding to 5. 
The rating of checkbox is: $3.62$ for checkbox only, and $3.68$ for checkbox+image, scenario.
Similar to the SUS score, the rating of checkbox is independent of the solving scenario. 
The rating of image is appreciably lower, at $2.84$.

Moreover, comparing preference scores from Figure \ref{fig: checkbox only preference} and
Figure \ref{fig: checkbox+image preference} with SUS scores from Table \ref{tab: reCAPTCHA sus score} 
we observe a trend for both checkbox and image: checkbox is more usable and rated positively, while
image is less usable and rated negatively. This leads to an unsurprising conclusion that participants'
preference for a given \rvii type is correlated with its usability level.

\begin{figure}
    \includegraphics[width=0.5\textwidth]{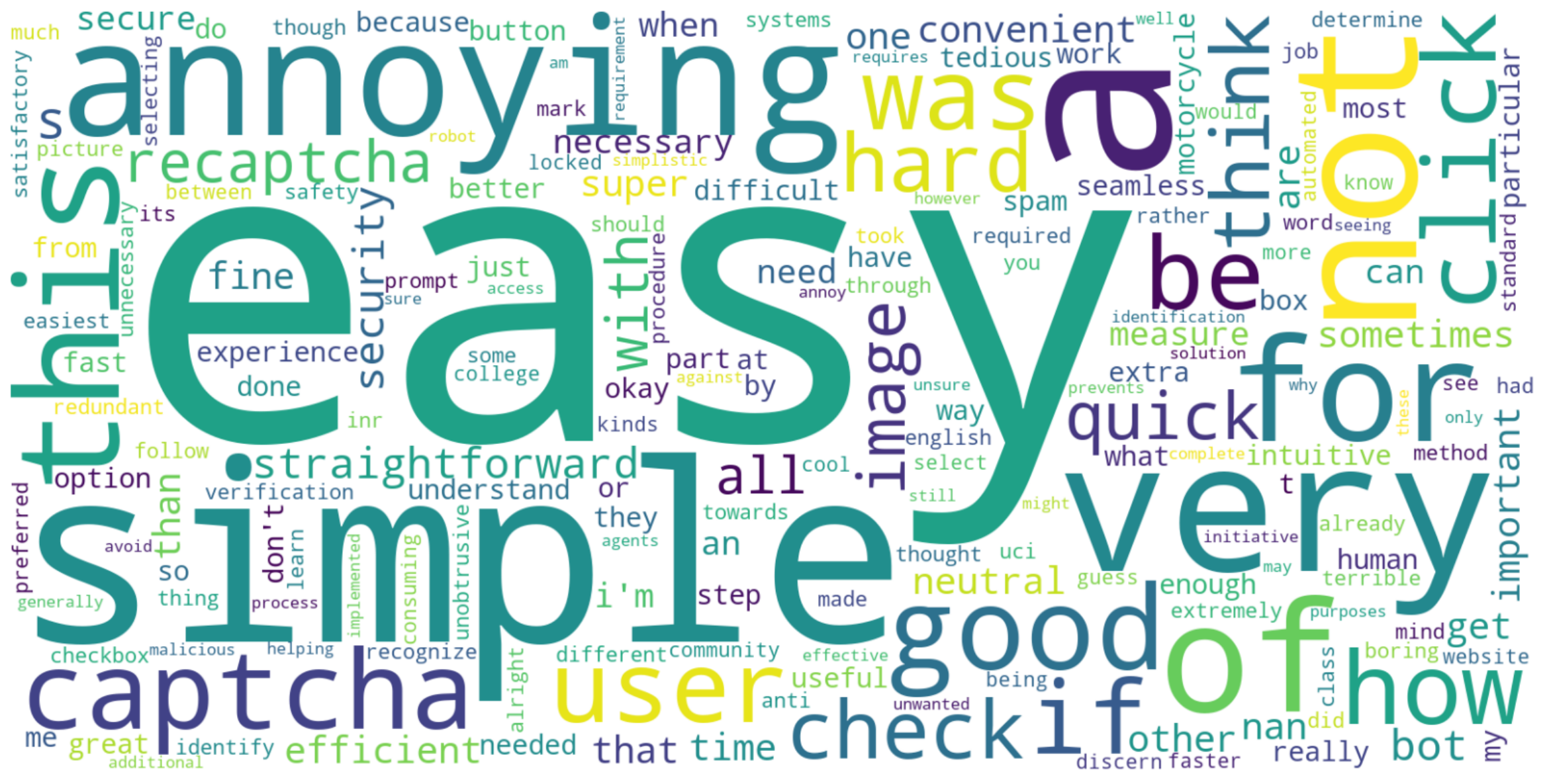}
    \caption{Word cloud from feedback on checkbox}
    \label{fig: word cloud checkbox}
\end{figure}

\begin{figure}
    \includegraphics[width=0.5\textwidth]{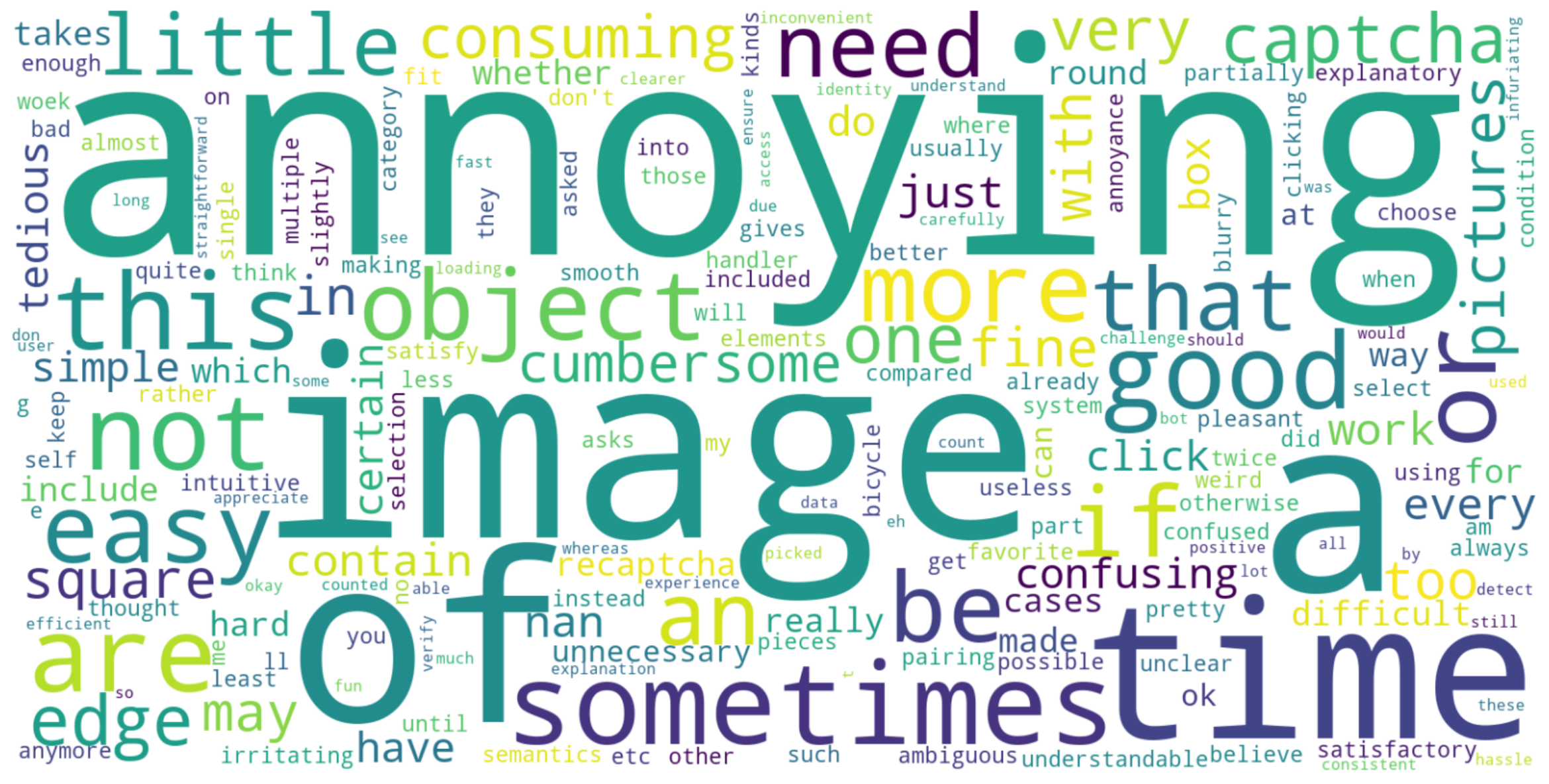}
    \caption{ Word cloud from feedback on image} 
    \label{fig: word cloud image}
\end{figure}

\subsubsection{Qualitative Feedback}
In the final part of the survey, participants were asked to provide open-ended feedback about 
checkbox and image using at least one word. Using collected feedback, we generate word clouds for both. 

The most prominent words for checkbox challenges in Figure \ref{fig: word cloud checkbox} are 
``easy'' and ``simple''. Other significant positive words are ``good'' and ``quick''. 
Nevertheless, checkbox is still labeled ``hard'' and ``annoying'' by some participants.

Figure \ref{fig: word cloud image} shows that the most prominent word describing image is
``annoying'', while a small fraction of participants label it as ``good'' and ``easy''. 
We acknowledge that the custom scale used in the survey might possibly introduce bias toward 
the word ``annoying''. However, the scale also includes the word ``pleasant'' and that word 
is not present in the word cloud as a positive opinion. Instead, participants used other positive-sounding 
words, such as ``easy'' or``simple''.

Negative words from the checkbox cloud and positive ones from the image cloud indicate that neither 
\rvii type is universally liked or disliked.
	\section{Comparison with Related Work}
\label{sec:related_work}
Prior results include \cite{Bursztein, Bigham, Gao, Ross, Uzun, manarDynamic2014, mohitAutomatic2019, gaoEmerging2019, 
Krol, fidas2011, senCAP, Ho, tanth19, searlesa_usenix}. They present average solving times for 
various \captchas, ranging from $3.1$ to $47$ seconds.
Compared with our observed mean solving time of $1.8$ seconds for checkbox \rvii, previous
results are $1.7$ to $26$ times slower. For image \rvii, our mean solving time is $10$ seconds, 
which is $3.3$ times slower than the fastest, and $5$ times faster than the slowest,
previously reported results. The fast solving time may be related to the trend noted in 
\cite{searlesa_usenix,Bursztein} of age influencing solving time: younger participants seem to solve 
faster than older ones. Since our population is mostly university students (aged $18-25$),
our results re-confirm this trend.

\begin{table}[!ht]
\footnotesize
\caption{Comparison with results from prior user studies evaluating \rvii: checkbox (C), image (I), total (T). Mean in seconds}
\label{tab:related_results}
\begin{tabularx}{\columnwidth}{l l X X X}
\toprule
Study                     & \textbf{Unique users}   & \textbf{\rviis  solved} & \textbf{Mean}           & \textbf{Accuracy} \\ \midrule
Ours                      & 3,625              &   9,141         &  10.4~(I), 1.85~(C), 3.53~(T)         &  93\%~(I), 80\%~(C) \\ \midrule
\cite{searlesa_usenix}    & 1,400               &   2,800         &  14-26~(I), 3.1-4.9~(C)               &  71-81\%~(I), 71-85\%~(C) \\ \midrule
\cite{tanth19}            & 40                  &  40            &  3.1 (T)  & None \\ \bottomrule
\end{tabularx}
\end{table}

To the best of our knowledge, only two prior efforts studied \rvii: \cite{searlesa_usenix} and \cite{tanth19}.
Table~\ref{tab:related_results} shows a direct comparison of the results.
However, \cite{tanth19} provides a very limited data set of ($n=40$) \rvii, containing only total times.
Whereas, \cite{searlesa_usenix}, provides the following points of comparison:
\begin{compactenum} 
    \item Amazon Mturk {\it vs} ``real world'' participants
    \item Participant awareness {\it vs} unawareness of the study existence and purpose
    \item Mock {\it vs} real account creation; 
    \item Preferences/Rating
\end{compactenum} 
Webb et al.~\cite{webb2022too} reported several points of concern about the quality of data collected from 
MTurk~\cite{mturk}. Our data and results are derived from a real-world scenario of actual users creating real accounts 
for a real service. 
However, since both this work and \cite{searlesa_usenix} implement \rvii in a similar way, some interesting conclusions can be 
drawn regarding the efficacy of Mturk data. Mturk users in \cite{searlesa_usenix} solved easy checkbox challenges $1.7-2.7$
times slower than our participants. They also solved easy image challenges $1.6-2.6$ times slower than our participants.
Another consideration is network speed, since MTurkers were participating in the \cite{searlesa_usenix} study
over the Internet. In contrast, our study was conducted with most\footnote{Recall that VPN use was required to 
create an account or recover a password, thus taking part in our study. Although most participants were on campus, 
some were remote. The exact number of the latter is unknown.}
participants being in close network proximity. Therefore, it would explaing why Mturk results are slower since
they can originate anywhere in the world, according to demographics reported in \cite{searlesa_usenix}.
This may also skew our results to be faster than the actual total \rvii solving time.

\cite{searlesa_usenix} showed that participants' awareness of the true purpose of the study could alter solving 
times. The solving time of participants who thought that they were participating in an account creation study was up to 57.5\% slower than those who knew that they were participating in a study about solving \captchas.
Account creation solving times (in seconds) for easy \rvii of \cite{searlesa_usenix} are $4.9$ for checkbox, 
and $26.3$ for image. In contrast, our results are $2.02$ for checkbox and $10.5$ for image (on the first attempt).
This translates into an average of $2.5$ times slowdown for both challenge types.
On the first attempt, our participants show the slowest mean and the least awareness (no study information presented) 
for checkbox challenges across significant groups ($n=2,888$), and our results show that solving time improves with 
subsequent attempts. Whereas, in \cite{searlesa_usenix} participants solved $10$ \captchas (among them $2$ \rviis) 
in sequence, which could lower the timing due to the repeated attempts bias. 

Also, \cite{searlesa_usenix} observed a lot of task abandonment, which might be due to 
the mocked-up (fake) account creation in that study. This is unlike our case where participants must create accounts
due to the SICS school-wide policy. Thus, they must complete the form with successful post-validation by the back-end server.
(In other words, abandonment is not an option).

\cite{searlesa_usenix} did not validate form information during account creation form submission beyond checking form field constraints, which could significantly alter the user study experience.
Since our high average multiple attempts per participant of 3.52 for checkbox and 1.73 for image was likely due to failed post-validation by the back-end server.

Study participants in \cite{searlesa_usenix} rated \rvii on a Likert scale, from ``least enjoyable'' to ``most enjoyable''.
Results showed that checkbox was the most enjoyable out of all \captchas, while image challenges were the least so.
The term ``enjoyable'' is synonymous with pleasant (the opposite of ``annoying''), which presents a point of comparison.
Our results in Figures~\ref{fig: checkbox only preference} and~\ref{fig: checkbox+image preference} are very 
similar in terms of positive and negative responses, thereby confirming results of \cite{searlesa_usenix}

	\section{Discussion}
\label{sec:discussion}

\subsection{Cost Analysis}
\label{sec:cost}
We now attempt to quantify various costs incurred by global use of \rv on the internet.
In particular we want to estimate the total time spent solving \rvs, the overall amount of human labor, 
network traffic (bandwidth), power consumption and the consequent environmental impact.
Note that, in the informal analysis below, we consider all estimates to be generous lower bounds.

Given that historic average solving time for distorted-text \captchas (same type used by \rvi) was 9.8 seconds and the conservative rate of 
100 million \rvs per day~\cite{recap_FAQ_100mil}, 980 million seconds per day were spent solving \rvis.
For \rvi, it lived from 2009-2014 for 5 years amounting to 183 billion \rvi sessions, taking 1.79 trillion seconds, or 497 million hours of human time spent solving \rvi. 
Given that the US federal minimum wage is \$7.5, this roughly yields \$3.7 billion in free wages.

Given that average solving time for all \rvii sessions is 3.53 seconds and the conservative rate of 
100 million \rvs per day~\cite{recap_FAQ_100mil}, 353 million seconds per day are spent solving \rviis.
For \rvii, it has been 9 years amounting to 329 billion \rvii sessions taking 1.16 trillion seconds, or 322 million hours of human time spent solving \rvii. 
Given that the US federal minimum wage is \$7.5, this roughly yields \$2.4 billion in free wages.

Assuming un-cached scenarios from our technical analysis (see Appendix~\ref{sec:technical_analysis}), 
network bandwidth overhead is 408 KB per session. This translates into 134 trillion KB or 134 Petabytes 
(194 x 1024 Terrabytes) of bandwidth.
A recent (2017) survey \cite{data_cost} estimated that the cost of energy for network data transmission was 
0.06 kWh/GB (Kilowatt hours per Gigabyte). Based on this rate, we estimate that 7.5 million kWh of energy 
was used on just the network transmission of \rv data. This does not include client or server related energy costs.
Based on the rates provided by the US Environmental Protection Agency (EPA) \cite{EPA_CALC} and US Energy Information 
Administration (EIA) \cite{EIA_FAQ}, 1 kWh roughly equals 1-2.4 pounds of CO2 pollution. This implies that
\rv bandwidth consumption alone produced in the range of 7.5-18 million pounds of CO2 pollution over 9 years.

In total from \rvi and \rvii: There have been at least 512  billion \rv sessions taking 2.95 trillion seconds, or 819 million hours, which is at least \$6.1 billion USD in free wages.
Out of the 329 billion \rvii sessions, (Our rate of 20\%) at least 65.8 billion would have been image challenges, while 263.2 million would have been checkbox challenges.
Thus 250 billion challenges would have resulted in labeled data.
According to Google, the value of $1,000$ items of labeled data is in the \$35-129 USD range~\cite{google_pricing}, which would be worth at least \$8.75-32.3 billion USD per each sale.

Lastly, we look into the economics of tracking cookies, another main by-product of \rv. 
Tracking cookies play an ever-increasing role in the rapidly growing online advertisement market. According to Forbes \cite{ForbesOnlineAd}, digital ad spending reached over \$491 billion globally in 2021, and more than half of the market (51\%) heavily relied on third-party cookies for advertisement strategies \cite{cookieReliance}. 
The expenditure on third-party audience data (collected using tracking cookies) in the United States reached from \$15.9 billion in 2017 to \$22 billion in 2021 \cite{expenseThirdParty}. 
More concretely, the current average value life-time of a cookie is €2.52 or \$2.7 \cite{miller2023economic}.
Given that there have been at least 329 billion \rvii sessions, which created tracking cookies, that would put the estimated value of those cookies at \$888 billion dollars.



\subsection{Security Analysis}
In the following subsections~\ref{sec:rv2} and \ref{sec:rv3}, we discuss different attacks that have been performed successfully against \rv. We consider behavior-based, image, and audio challenges in \rvii and \rviii. 
Table~\ref{tab:bots_vs_humans} shows a direct comparison of the time and accuracy for humans and bots.

\begin{table}[h!]
\centering
\footnotesize
\caption{Humans vs.\ bot solving time (seconds) and accuracy (percentage) for \rvii.}
\label{tab:bots_vs_humans}
\begin{tabularx}{\columnwidth}{l l l l l l l}
\toprule
      &       \multicolumn{4}{c}{\textbf{Human}}                    & \multicolumn{2}{c}{\textbf{Bot}}  \\  \cmidrule(lr){2-5}\cmidrule(lr){6-7} 
\textbf{Type}  &  Time         &   Acc    &  Time          &   Acc      &  Time                        &  Acc                    \\ \midrule
checkbox       &   1.85     &   80\%     &   3.1-4.9  \cite{searlesa_usenix}     &  85\%   \cite{searlesa_usenix}       & 1.4 \cite{Sivakorn2016}      & 100\% \cite{Sivakorn2016}    \\
image       &  10.4        &  93\%         &  16-26   \cite{searlesa_usenix}       &  81\%   \cite{searlesa_usenix}          & 17.5  \cite{Hossen2020}      & 85\% \cite{Hossen2020}       \\ 
\bottomrule
\end{tabularx}
\end{table}

\subsection{\rvii}
\label{sec:rv2}
\rvii presents three different types of captcha challenges to the users: behavior-based (checkbox) challenge, image challenge, and audio challenge. Unfortunately, each of these captcha types has been proven vulnerable to attacks.

\subsubsection{Checkbox Challenge} 
With the introduction of \rvii, came a new serious vulnerability in the form of click-jacking \cite{the_NOCAPTCHA_problem}.
Adversaries can make "trustworthy" users generate g-recaptcha-response-s, which can be automatically used to pass challenges, ultimately making a Bot's job infinitely easier!

Sivakorn, et al. \cite{Sivakorn2016} perform an in-depth analysis of the risk analysis system of \rv 
and implement an attack to manipulate it. Based on this analysis and implementation:
\begin{compactenum} {
    \item Google primarily uses tracking cookies in the risk analysis system.
    \item At least 63,000 valid cookies can be automatically created per day per IP address.
    \item 9 days after a cookie creation, checkbox attempts using the cookie will succeed.
    \item 52,000-59,000 checkbox challenges can be solved with 100\% accuracy per day per IP address.
    \item The average solution time is 1.4 seconds with 100\% accuracy, shown in Table~\ref{tab:bots_vs_humans}.
    }
\end{compactenum} 

Given the blatant vulnerability \cite{the_NOCAPTCHA_problem}, ease of implementing large-scale automation \cite{Sivakorn2016}, and usage of privacy invasive tracking cookies \rvii checkbox presents itself as a complete vulnerability disguised as a security tool.
Google was previously sued 22.5 million for \textbf{secretly} adding tracking cookies to apple users devices \cite{Ritchie_Jayanti_2019}.
It can be concluded that the true purpose of \rvii is as a tracking cookie farm for advertising profit masquerading as a security service.

\subsubsection{Image Challenge}
Image-labeling challenges have been around since 2004 with the introduction of Image Recognition \captchas by Chew et al.~\cite{image_captchas}.
6 years later, in 2010 Fritsch et al.~\cite{attacking_images} published an attack that beat the prevalent image \captchas of the time with 100\% accuracy.
At this point, it could be concluded that image recognition was no longer difficult to solve automatically with a computer.
However in 2014 with the introduction of \rvii, the fall-back security method was an image challenge, which had been proven insecure 4 years prior.
The idea is that if your cookies aren't valuable enough then \rvii would present an image labeling task.
This wouldn't make sense as a security service, yet it would make sense given that obtaining labeled image data is highly valuable and is even sold by Google~\cite{google_pricing}.
The conclusion can be extended that the true purpose of \rvii is a free image-labeling labor and tracking cookie farm for advertising and data profit masquerading as a security service.

Consequently, \cite{Sivakorn2016} and \cite{Hossen2020} investigate and successfully implement automated solutions to \rviis image labeling task.
In 2016, \cite{Sivakorn2016} showed that a plethora of automated services, 
including Google's own Google Reverse Image Search (GRIS), 
could be used to automatically complete \rviis image labeling tasks.
\cite{Sivakorn2016} also implemented its own easy solver with 70.8\% accuracy at 19.2 seconds per \rvii image labeling task.
In 2020, \cite{Hossen2020} also showed that many automated services, 
including Google's own Google Cloud Vision, 
could be used to automatically complete \rviis image labeling tasks with reasonable speed and accuracy.
\cite{Hossen2020} similarly implemented an attack, achieving a high level of speed (17.5 seconds) and accuracy (85\%), shown in Table~\ref{tab:bots_vs_humans}.

\subsubsection{Audio Challenge}
As part of \rvii, Google introduced accessibility options allowing users to use audio \captchas, instead of image-based ones.
Unsurprisingly, these audio \captchas introduce an accessibility side-channel, especially apparent due to advances in 
speech-to-text technology.

In 2017, Bock, et al. \cite{unCaptcha} introduced an automated system called {\it unCaptcha} which can solve 
audio challenges with $85.15\%$ accuracy and $5.42$ seconds average solving time. 
Similar to other attacks \cite{unCaptcha} uses Google's own voice recognition technology 
as a means to break audio challenges.

\subsection{\rviii}
\label{sec:rv3}
\rviii was introduced in 2018 \cite{intro_reCAPTCHAv3} proposing the returning of a score, which website developers could use 
to decide whether to prompt with a challenge or perform some other action.
Challenges types served by \rviii are the same as \rvii. Also, there is no discernible difference between \rvii and \rviii
in terms of appearance or perception of image challenges and audio challenges.
Hence, attacks targeting \rvii image/audio challenges are also applicable for those of \rviii. 
However, assuming that the risk analysis system was updated from \rvii to \rviii, breaking behavior-based 
challenges of \rviii might require new techniques.
In 2019, Akrou, et al. \cite{akrout2019hacking} presented a reinforcement learning (RL) based 
attack breaking \rviii's behavior-based challenges, obtaining high scores ($.9+$), with $97\%$ 
accuracy and only requiring $2,000$ data points as a training set.

	\section{Summary} 
\label{sec:conclusion}
Over 13 years passed since \rvs initial appearance and its current prevalence is undeniable.
It is thus both timely and important to investigate its usability. 
This paper presents a real-world user study with over $3,600$ unbiased (unwitting) participants 
solving over $9,000$ \rvii challenges. We explore four new dimensions of \rvii solving time: \# of 
attempts, service type, as well as educational level and major. Results show that: 
\begin{compactitem}    
    \item Participants improve in terms of solving time with more attempts, for checkbox challenges.
    \item The service/website setting is an important consideration for researchers and web developers, since 
    it has a statistically significant effect on solving time.
    \item Educational level directly impacts solving time.
    \item There were minor trends with statistical significance of participants with technical (STEM) majors 
    solving time being faster than that of others.
\end{compactitem}

In terms of usability, the post-study survey results show that the checkbox challenge gets an average SUS score of $77$. 
This is considered to be acceptable and preferred by many participants over the image challenge, which has an average
SUS score of $59$. Notably, participants found the image challenge to be annoying.

In terms of cost, we estimate that -- during over 13 years of its deployment --  $819$ million hours of 
human time has been spent on \rv, which corresponds to at least $\$6.1$ billion USD in wages.
Traffic resulting from \rv consumed $134$ Petabytes of bandwidth, 
which translates into about $7.5$ million kWhs of energy, corresponding to $7.5$ million pounds of CO2.
In addition, Google has potentially profited \$888 billion USD from cookies and \$8.75-32.3 billion USD per each sale of their total labeled data set.

In terms of security \rvii presents:
\begin{compactitem}    
    \item Click-jacking (a blatant vulnerability) \cite{the_NOCAPTCHA_problem}
    \item Trivial implementation of large-scale automation attacks \cite{Sivakorn2016}
    \item Weakness of security premise of fallback (image challenge)~\cite{attacking_images, Sivakorn2016, Hossen2020}
    \item Usage of privacy invasive tracking cookies (for security) \cite{Sivakorn2016}
\end{compactitem}
Ultimately, given these points it can be concluded that \rvii presents no real security.

Given that: (1) \rvii is negatively perceived by most users, (2) its immense cost,  and (3) its susceptibility to bots, 
our results prompt a natural conclusion: 
\noindent\begin{quote} {\it
\rvii and similar \rv technology should be deprecated.}
\end{quote}
	
	\ifdefined\showauthors
	\section{Acknowledgements}
	\fi

	\footnotesize
	\raggedbottom

    \bibliographystyle{ACM-Reference-Format}
    \bibliography{references.bib}

	\normalsize
	\appendix



\section{Workflow}
\label{sec:workflow}
In this appendix we show basic workflows for account creation and password recovery processes that participants 
followed in the user study. 

\subsection{Account Creation}
\begin{figure}[!ht]
    \centering
    \includegraphics[width=\columnwidth]
    {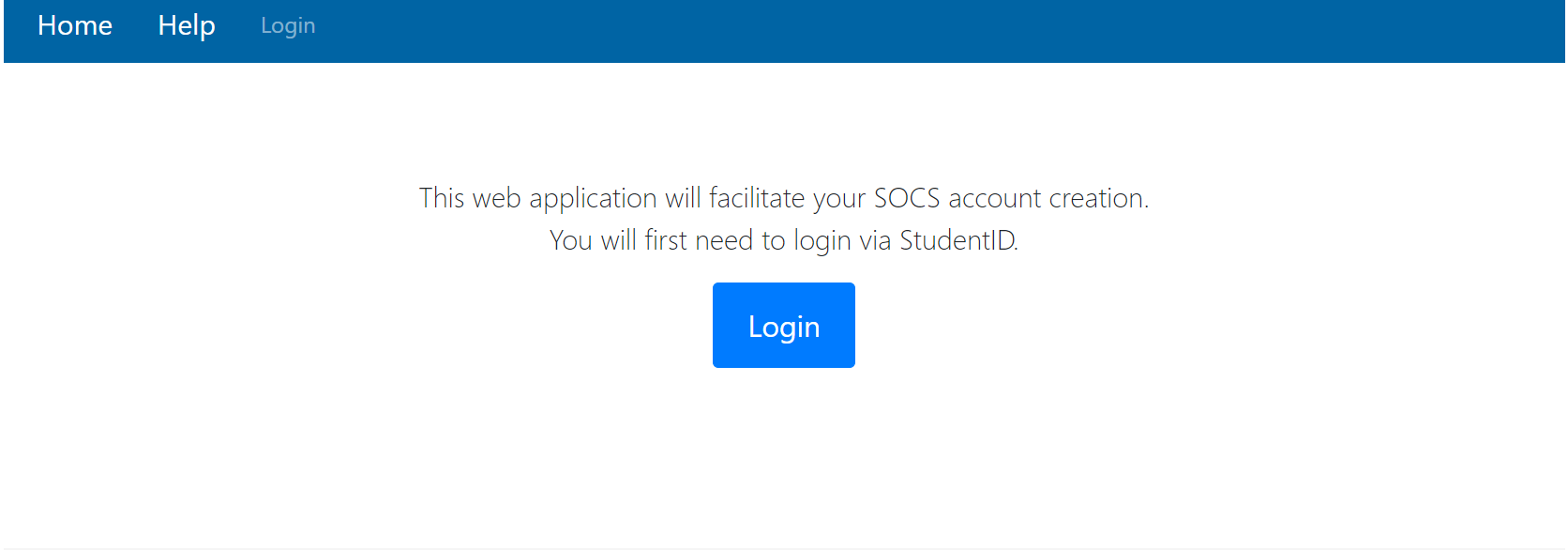}
    \caption{Initial login page}
    \label{fig:initial_login}
\end{figure}

\begin{figure}[!ht]
    \centering
    \includegraphics[width=\columnwidth]
    {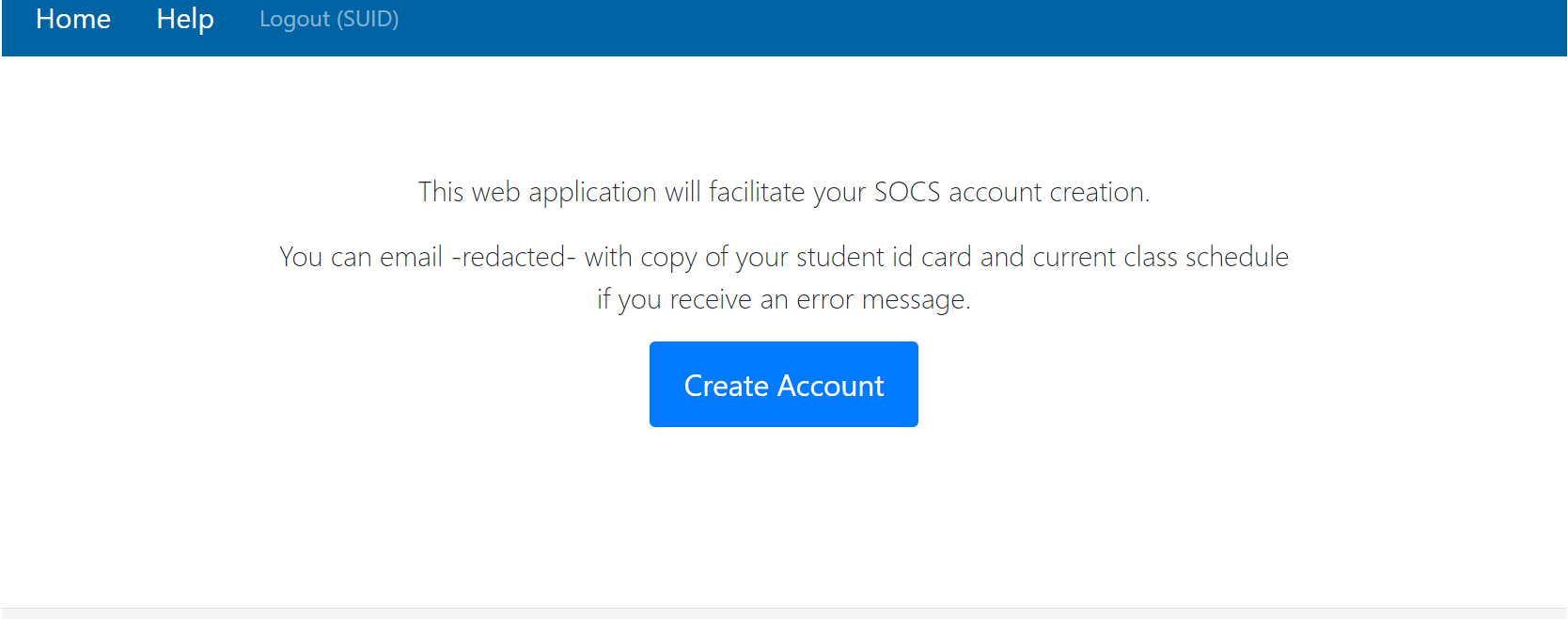}
    \caption{Initial Account Creation Page}
    \label{fig:initial_create_acc}
\end{figure}

\begin{figure}[!ht]
    \centering
    \includegraphics[width=\columnwidth]
    {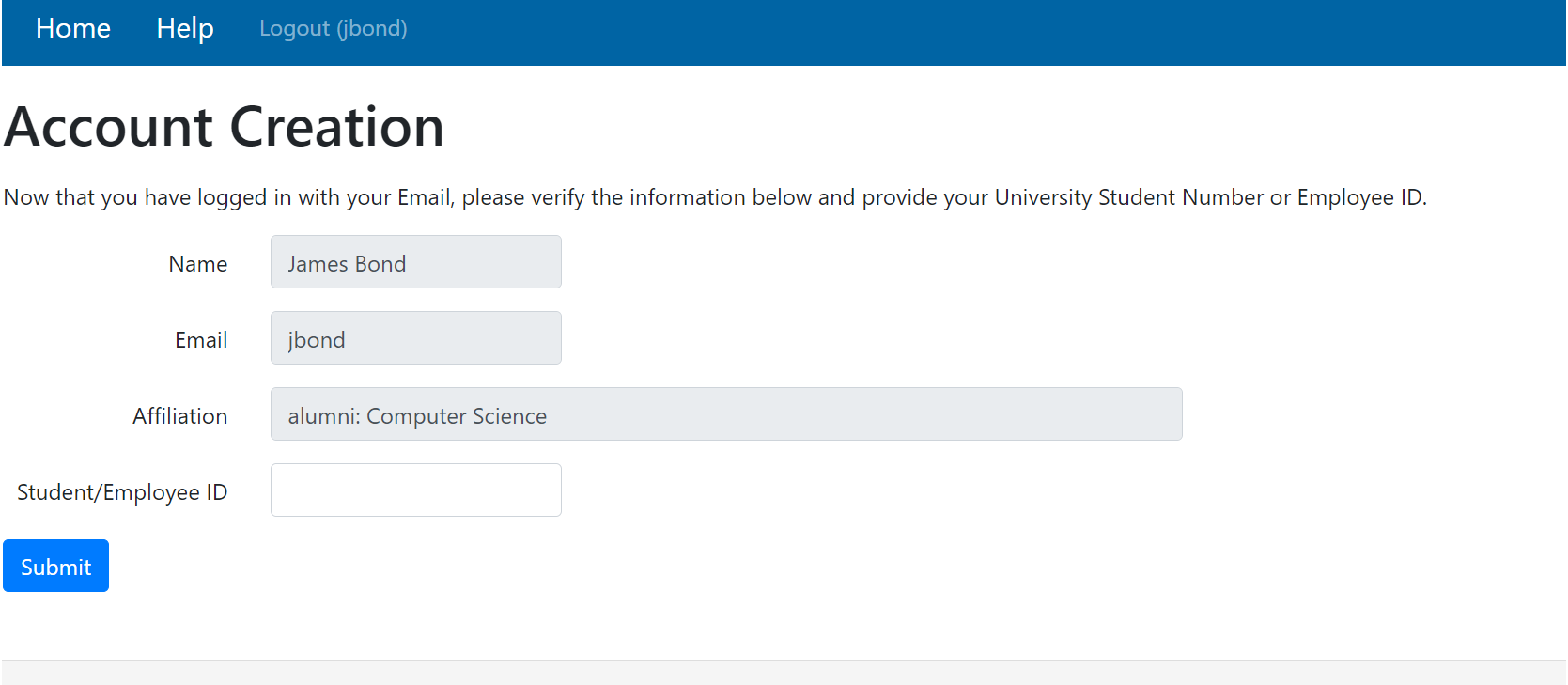}
    \caption{Account creation form}
    \label{fig:ac_pre_submit}
\end{figure}

\begin{figure}[!ht]
    \centering
    \includegraphics[width=\columnwidth]
    {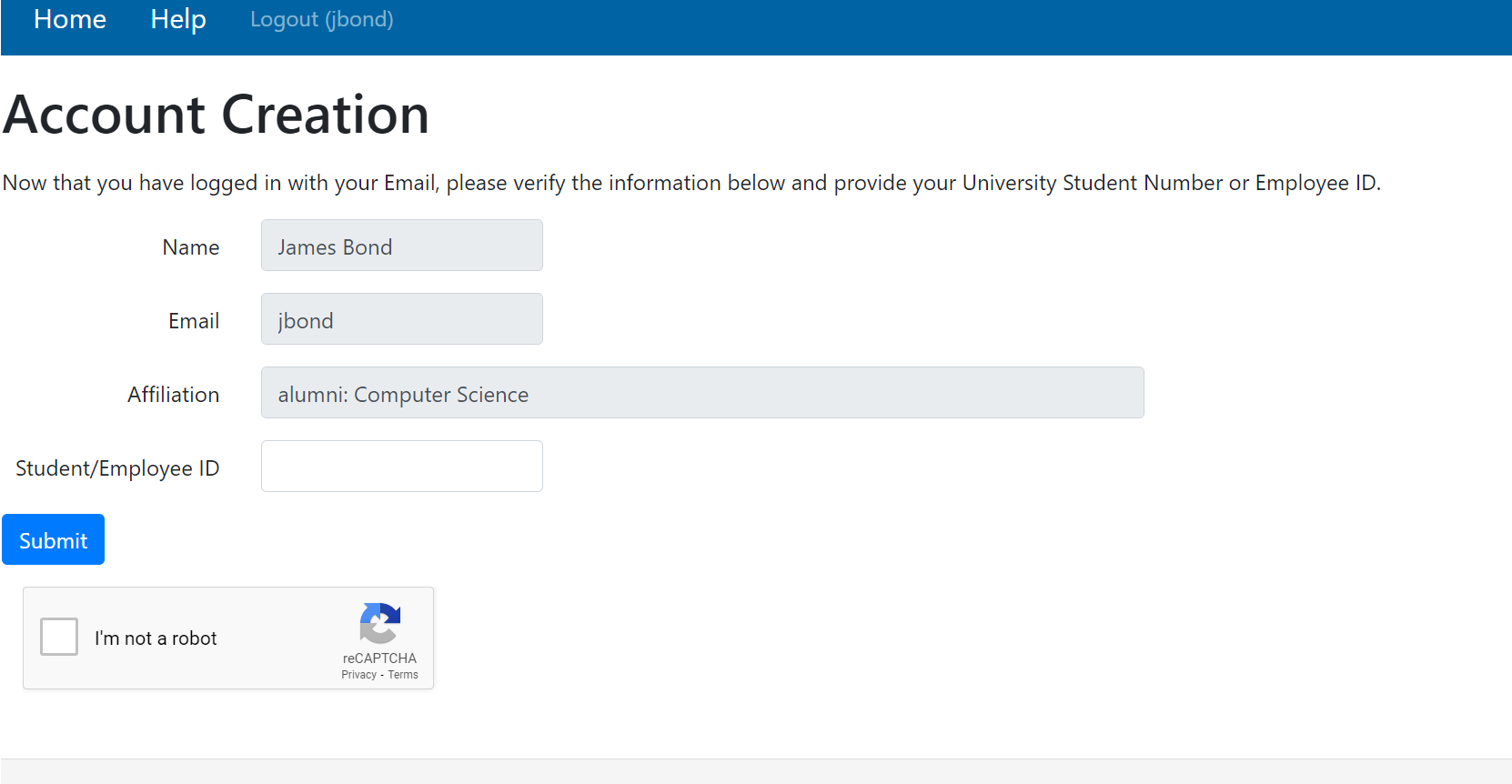}
    \caption{AC form after clicking submit}
    \label{fig:ac_post_submit}
\end{figure}

Figures \ref{fig:initial_login}, \ref{fig:initial_create_acc}, \ref{fig:ac_pre_submit}, \ref{fig:ac_post_submit} constitute the workflow of the account creation process.

\subsection{Password Recovery}
\begin{figure}[!ht]
    \centering
    \includegraphics[width=\columnwidth]
    {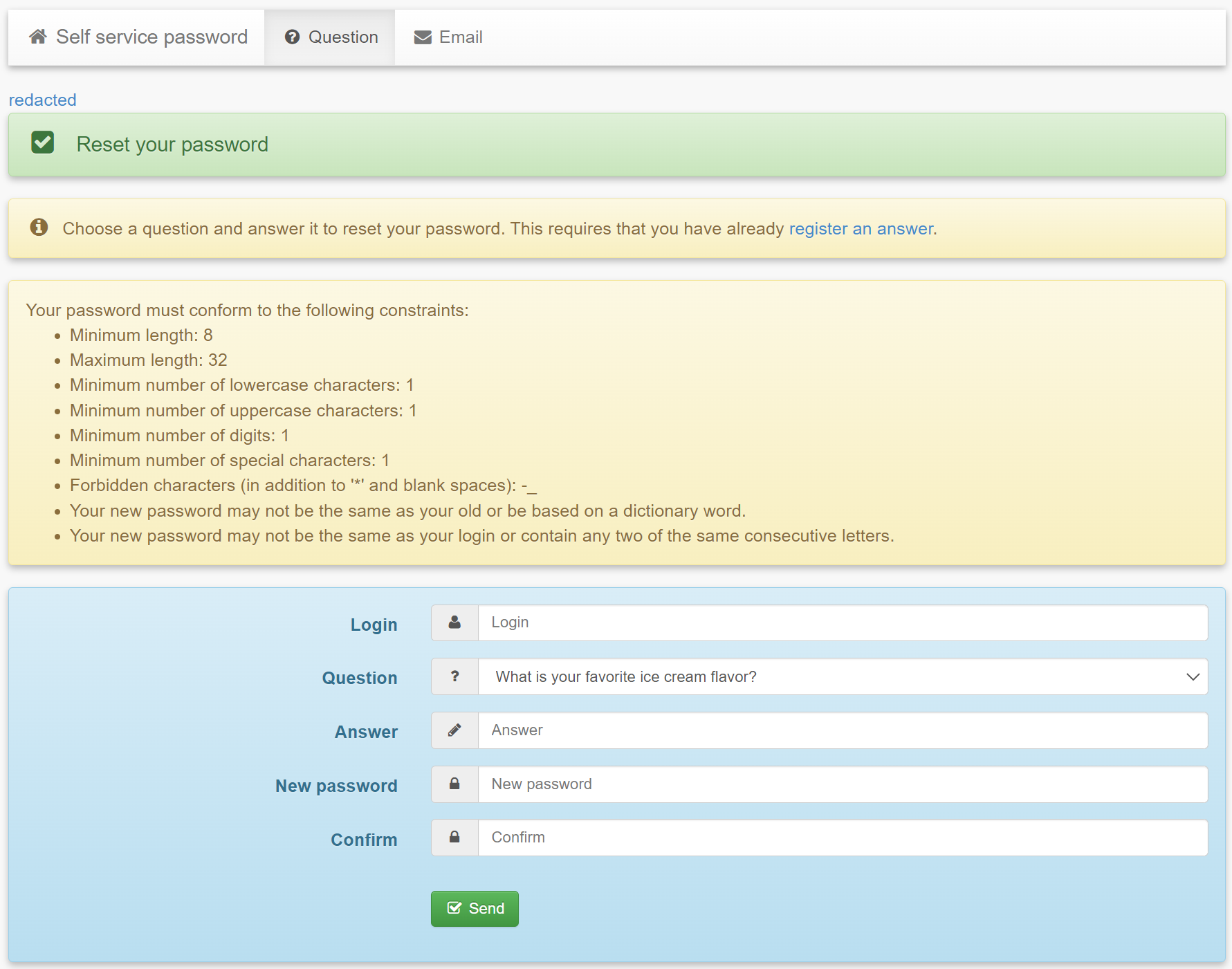}
    \caption{Password Recovery form}
    \label{fig:reset_pre_submit}
\end{figure}

\begin{figure}[!ht]
    \centering
    \includegraphics[width=\columnwidth]
    {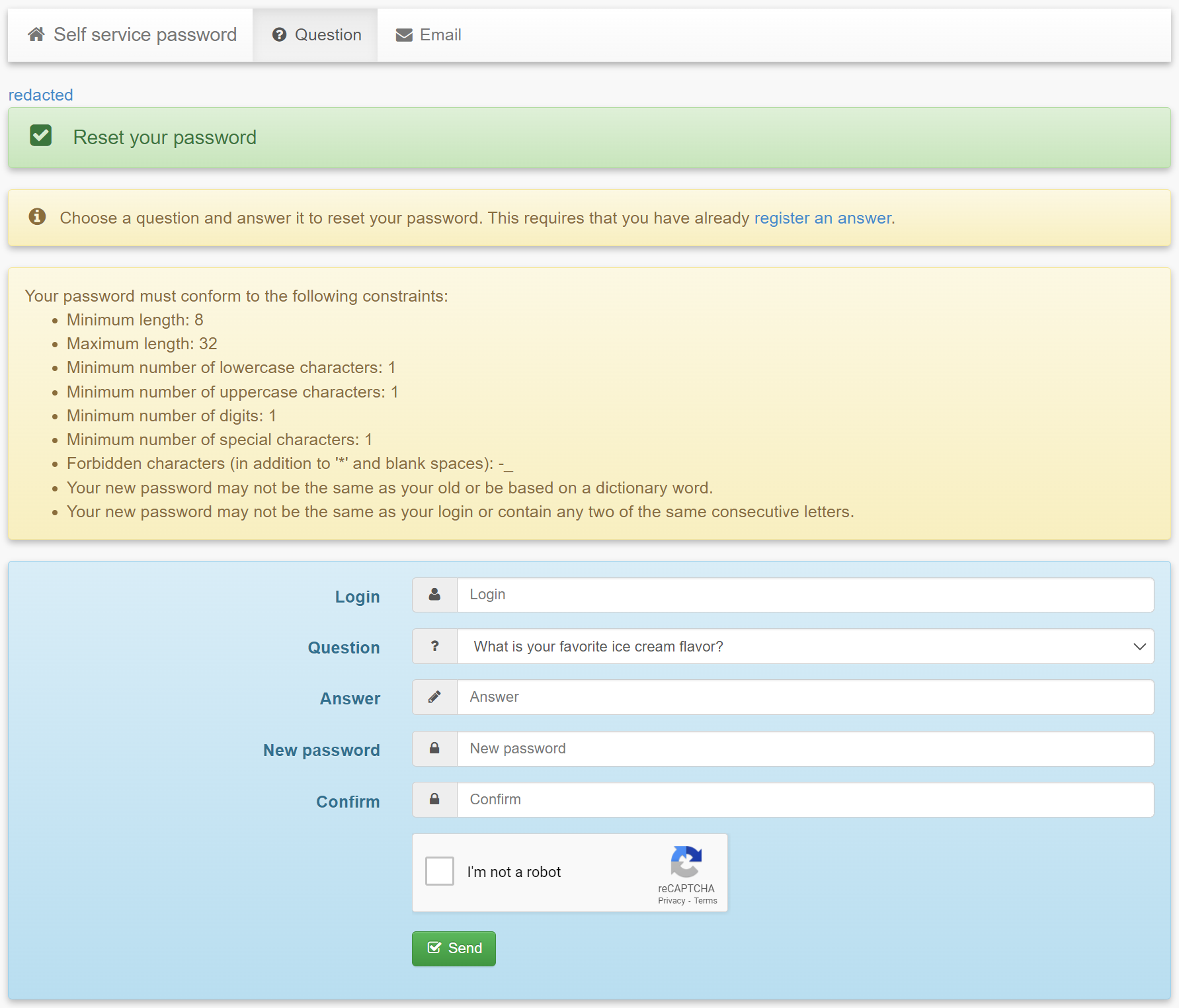}
    \caption{Password Recovery form after clicking submit}
    \label{fig:reset_post_submit}
\end{figure}

Figures \ref{fig:reset_pre_submit} and \ref{fig:reset_post_submit} present the workflow of the password recovery process.

\section{Network Analysis of {\rvii}}
\label{sec:technical_analysis}
\begin{figure}
\begin{mdframed} \footnotesize
  \begin{lstlisting}[language=html]
<html>
<head> <title> Simple Web Page </title> </head>
<body> <h4> A minimal web page </h4> <br/> </body>
</html>
\end{lstlisting}  
\end{mdframed}
\caption{Source code of simple.html}
\label{fig: simple page}
\end{figure}

\footnotesize
\begin{figure} 
\begin{mdframed} \footnotesize
\begin{lstlisting}[language=html] 
<html>
<head> <title> reCAPTCHA Difficult </title>
<script src="https://www.google.com/recaptcha/api.js" 
    async defer></script>
</head>
<body>
<h4>A minimal web page</h4>
<div class="g-recaptcha" 
    data-sitekey="obtained-site-key"></div>  <br/>
</body>
</html>
 \end{lstlisting}
\end{mdframed}
    \caption{Source code of recaptcha.html}
    \label{fig: recaptcha page}
\end{figure}
\normalsize

This Appendix contains a high-level technical analysis of \rv. It has been
considered in \cite{Sivakorn2016}, which described the display method and 
workflow of \rv with the emphasis on security aspects. Whereas, our goal is to: (1) determine various overhead factors 
incurred whenever a web-page uses \rv, and (2) investigate \rv's automation detection capability.
To this end, we performed black box program and network traffic analyses for common usage scenarios. We used two
simple web pages for this purpose:
\begin{itemize}
\item Baseline page without any \captchas. This page is called {\em simple.html} and its source code 
is shown in Figure \ref{fig: simple page}.
\item A page similar to the baseline page, except with an additional \rv. This page is called recaptcha.html 
and its source code is shown in Figure \ref{fig: recaptcha page}. As evident from the figure, 
integrating \rv into a web page is very easy and straightforward. 
\end{itemize}
These pages were visited using Google Chrome browser  \cite{chrome} and each usage scenario was repeated at least ten times. 
Browsing was performed in both guest and normal (profile logged-in) modes. Relevant information in the format of a {\em .har} 
file was collected for each scenario using Chrome DevTools. 

The rest of this section describes the findings. Notations are summarized in Table \ref{tab: notations}.

\begin{table}[!ht] \small
    \caption{Notation Summary}
    \begin{tabularx}{0.5\textwidth}{lX}
    \toprule
    Notation & Description    \\ 
    \midrule
    g1 & https://www.google.com/recaptcha   \\
    g2 & https://www.gstatic.com/recaptcha/releases/vkGiR-M4noX1963Xi\_DB0JeI\\
    g3 & https://www.gstatic.com/recaptcha/api2/ \\
    g4 & https://www.google.com/recaptcha/api2/ \\
    g5 & https://fonts.gstatic.com/s/roboto/v18/    \\
    dv & different values \\
    \bottomrule
    \end{tabularx}
    \label{tab: notations}
\end{table}

\subsection{Page load Latency}\label{sec:load}
Table \ref{tab: page load overhead} shows additional API calls made while loading {\em recaptcha.html} webpage.

\begin{table}[!ht] \small
\caption{ \rv API Calls during page load}
\begin{tabularx}{0.5\textwidth}{l l}
\toprule
Request URL & Content-Length (B) \\
\midrule
g1/api.js & 554 \\

g2/recaptcha\_\_en.js & 166822 \\

g4/anchor?ar=[dv] & 27864 (average) \\

g2/styles\_\_ltr.css & 24605 \\

g2/recaptcha\_\_en.js & 166822 \\

g3/logo\_48.png
 & 2228 \\

 g4/webworker.js?hl=[dv] & 112 \\

g2/recaptcha\_\_en.js
 & 166822 \\
 
 g4/bframe?hl=[dv] {\&}v=[dv]{\&}k=[dv] & 1141-1145 \\

g2/styles\_\_ltr.css & 24605 \\

g2/recaptcha\_\_en.js & 166822 \\
\midrule
Network Overhead & 254.01 KB-316.64KB\\
\bottomrule
\end{tabularx}
\label{tab: page load overhead}
\end{table}

There are also 2-to-6 calls to g5 for downloading various web fonts. Content length for each of 
these calls is 15340, 15344, and 15552 bytes. Even though multiple calls are made to download 
recaptcha\_\_en.js and styles\_\_ltr.css, only the first call downloads the file, if necessary. 
These observations are taken into account when computing network overhead in Table \ref{tab: page load overhead}.

Moreover, api.js, recaptcha\_\_en.js, styles\_\_ltr.css, logo\_48.png, and web fonts are often served from the 
cache. Table \ref{tab: page load overhead} provides an upper bound on network overhead for page load. 
Average network overhead is computed by extracting actual network transmission during page load from collected 
{\em .har} files. Table \ref{tab: actual network overhead} shows the results.

\begin{table}[!ht] \small
\caption{recaptcha.html load network overhead}
\begin{tabularx}{0.5\textwidth}{l l l}
\toprule
Scenario & Page Name & Page Size(KB) \\
\midrule
First load  & simple.html & 0.631KB\\
First load & recaptcha.html & 408.5KB\\
\midrule
 Network Overhead &  & 407.869KB\\
\midrule 
Subsequent loads & simple.html & 0.241 KB\\
Subsequent loads & recaptcha.html & 29.56 KB\\
\midrule
 Network overhead &  & 29.319 KB \\
\bottomrule
\end{tabularx}
\label{tab: actual network overhead}
\end{table}

We investigated load latency using Chrome DevTools, pingdom.com \cite{pingdom}, and webpagetest.com \cite{webpagetest}. Table 
\ref{tab: load time overhead} presents the results. Latency computed using Chrome DevTools is the highest 
since Chrome DevTools determines the load time of simple.html and recaptcha.html in the same network 
where the concerned web pages are hosted. Observation shows that load latency increases as the distance
between the user and the hosted webpage decreases (in terms of hops).

\begin{table}[!ht]\footnotesize
\caption{ recaptcha.html load latency}
\begin{tabularx}{0.5\textwidth}{l l l} 
\toprule
Measurement Tool & Page Name & load Time \\
\midrule
Chrome DevTools & simple.html & 51.16ms\\
Chrome DevTools & recaptcha.html & 425.81ms\\
\midrule
Time Overhead &  & 374.65ms, 732.31\%\\
\midrule 
pingdom.com  & simple.html & 375ms\\
pingdom.com & recaptcha.html & 796ms\\
\midrule
Time Overhead &  & 471ms, 125.6\%\\
\midrule 
webpagetest.org & simple.html & 814.22ms\\
Subsequent Loads & recaptcha.html & 2074.78ms\\
\midrule
 Latency &  & 1260.56ms, 154.82\% \\
\bottomrule
\end{tabularx}
\label{tab: load time overhead}
\end{table}

\subsection{Checkbox Click Overhead}\label{sec:sclick}
Table \ref{tab: checkbox click overhead} shows additional API calls made after checkbox is clicked. 
In this scenario, image \captcha is not served to the user.

\begin{table}[!ht] \small
\caption{\rv API Calls after checkbox click}
\begin{tabularx}{0.5\textwidth}{l l}
\toprule
Request URL & Content-Length (B) \\
\midrule
g4/reload?k=[dv] & 23844.67 (average) \\
g4/userverify?k=[dv] & 580.56 (average) \\
g3/refresh\_2x.png & 600 \\

g3/audio\_2x.png
 & 530 \\

g3/info\_2x.png
 & 665 \\
g5/[font].woff2 & 15552 \\
\midrule
Network Overhead & 24.43 KB-41.77KB\\
\bottomrule
\end{tabularx}
\label{tab: checkbox click overhead}
\end{table}

In some cases, only the first two calls are made. Even when other calls are made, files are normally served from 
the cache, so there is no network traffic. Files are downloaded only in the first-ever attempt to solve \rv in 
a given client browser.
Table \ref{tab: checkbox click overhead} depicts upper and lower bounds for the network overhead.

\subsection{\rv Image load Overhead}\label{sec:image}
Table \ref{tab: image load overhead} shows additional API calls made when checkbox is clicked and an image \captcha 
is loaded. It also provides the upper bound and the lower bound of the network overhead due to these calls.

\begin{table}[!ht] \small
\caption{\rv API Calls for image load}
\begin{tabularx}{\columnwidth}{l l}
\toprule
Request URL & Content-Length (B) \\
\midrule
g4/reload?k=[dv] & 24439.16667 (average) \\

g3/refresh\_2x.png
& 600 \\

g3/audio\_2x.png
 & 530 \\

g3/info\_2x.png
 & 665 \\
g4/payload?p=[dv] & 39589.45455 (average) \\
\midrule
Network Overhead & 64.03 KB-96.72KB\\
\bottomrule
\end{tabularx}

\label{tab: image load overhead}
\end{table}

In some cases, two calls are made to g5 to download web fonts; content length is 15340 and 15552 bytes, respectively. 
Also, refresh\_2x.png, audio\_2x.png, info\_2x.png, and web fonts are often served from the cache instead of being downloaded.

\subsection{Image Solution Verification Overhead}\label{sec:verification}
Table \ref{tab: image verification overhead} shows additional API calls made when an image \captcha solution is verified. 
In case of a correct solution, only the third call from Table \ref{tab: image verification overhead} requires network 
transmission and thus incurs network overhead. In case of a wrong solution, the last call from Table \ref{tab: image verification overhead} 
is made, which requires network transmission and adds to network overhead. 
In both cases, other calls are usually served from the cache.
In some instances, when a wrong solution occurs, only the third and 
fifth calls from Table \ref{tab: image verification overhead} are made.

\begin{table}[!ht] \small
\caption{\rv API Calls for correct image solution}
\begin{tabularx}{0.5\textwidth}{p{2.5cm}p{3cm}p{3cm}}
\toprule
Case & Request URL & Content-Length (B) \\
\midrule
Both  & g3/refresh\_2x.png & 600 \\
 
Both & g3/audio\_2x.png & 530 \\
 
Both & g4/userverify?k=[dv] & 595.88\\
  
Both & g3/info\_2x.png & 665 \\

Wrong Solution & g4/payload?p=[dv] & 40922.167 (average) \\
\midrule
\multicolumn{2}{c}{Correct Solution Network Overhead}  & 0.6KB\\
 \midrule
\multicolumn{2}{c}{Wrong Solution Network Overhead}  & 41.58KB\\
\bottomrule
\end{tabularx}
\label{tab: image verification overhead}
\end{table}

\begin{table}[!ht] \small
\caption{\rv API Calls for \rv expiration}
\begin{tabularx}{0.5\textwidth}{ll}
\toprule
Request URL & Content-Length (B) \\
\midrule
 g4/anchor?ar=[dv] & 27864 (average) \\

 g2/styles\_\_ltr.css & 24605 \\

 g2/recaptcha\_\_en.js & 166822 \\

 g3/logo\_48.png
 & 2228 \\

 g4/webworker.js?hl=[dv] & 112 \\

 g2/recaptcha\_\_en.js
 & 166822 \\
 
 g4/bframe?hl=[dv] {\&}v=[dv]{\&}k=[dv] & 1141-1145 \\

 g2/styles\_\_ltr.css & 24605 \\

 g2/recaptcha\_\_en.js & 166822 \\
\midrule
Network Overhead & 29KB\\
\bottomrule
\end{tabularx}

\label{tab: recaptcha expired overhead}
\end{table}

\subsection{\rv Expiration Overhead}\label{sec:expiration}
Table \ref{tab: recaptcha expired overhead} shows additional API calls made after a \rv solution expires.
Only the first and seventh calls (g4/anchor and g4/bframe) require network transmission and are considered in
network overhead. Other calls are served from the cache.

\noindent{\bf Summary:}
Results of evaluating network overhead for various \rv usage scenarios are summarized in 
Table \ref{tab: network overhead summary}. As evident from these results, using \rv incurs 
considerable network and timing overhead. 

\footnotesize
\begin{table}[!ht] 
\caption{Summary of \rv Network Overhead}
\label{tab: network overhead summary}
\begin{tabularx}{0.5\textwidth}{l l}
\toprule
Scenario & Network Overhead(KB) \\
\midrule
First time Page Load & 408.5 \\
Subsequent Page Loads & 29.319  \\
Checkbox Click & 24.43-41.77\\
Image Load & 64.03-96.72\\
Image Correct Solution Verification & 0.6\\
Image Wrong Solution Verification \& New Image load & 41.58\\
Solution Expiration & 29\\
\bottomrule
\end{tabularx}
\end{table}
\normalsize

\subsection{Automation Detection}\label{sec:bot detection}
Finally, we briefly looked into automation detection capability of \rv. Specifically, checkbox 
click is performed through Jitbit mouse macro recorder \cite{mouse_macro} and playwright automated headless 
Chrome browser \cite{playwright}. Interestingly, the use of the mouse macro is not considered as suspicious bot 
activity by \rv. When checkbox is clicked and the page is reloaded in quick succession, an 
image \captcha is served on around 14 tries, regardless of whether the tasks were performed manually or via the 
mouse macro. However, performing the same tasks via Playwright Chrome browser is considered suspicious -- an Image 
\captcha is served upon the first request.

	
	\normalsize
	
	\ifdefined\showchanges
	\onecolumn
	\twocolumn
	\fi
	
\end{document}